\documentclass[aps,prd,showpacs,showkeys,preprintnumbers,notitlepage,superscriptaddress]{revtex4-1}
\usepackage{amsmath,amssymb,amsthm,bm,bbm,color,hyperref}
\usepackage{graphicx,subfigure,rotating}

\begin{document} 

\title{Center Vortices, Area Law and the Catenary Solution}

\author{Roman H\"ollwieser\footnote{Funded by an Erwin Schr\"odinger Fellowship of the Austrian Science Fund under Contract No. J3425-N27.}}
\email{hroman@kph.tuwien.ac.at}
\affiliation{Department of Physics, New Mexico State University, PO Box 30001, Las Cruces, NM 88003-8001, USA}
\affiliation{Institute of Atomic and Subatomic Physics, Nuclear Physics Dept.\\ Vienna University of Technology, Operngasse 9, 1040 Vienna, Austria}
\author{Derar Altarawneh}
\email{derar@nmsu.edu}
\affiliation{Department of Physics, New Mexico State University, PO Box 30001, Las Cruces, NM 88003-8001, USA}
\affiliation {Department of Applied Physics, Tafila Technical University, Tafila , 66110 , Jordan}
\iffalse
\author{Michael Engelhardt}
\email{engel@nmsu.edu}
\affiliation{Department of Physics, New Mexico State University, PO Box 30001, Las Cruces, NM 88003-8001, USA}
\fi
\date{\today}

\begin{abstract}
We present meson-meson (Wilson loop) correlators in $Z(2)$ center vortex models for the infrared sector of Yang-Mills theory, {\it i.e.}, a hypercubic lattice model of random vortex surfaces and a continuous 2+1 dimensional model of random vortex lines. In particular we calculate quadratic and circular Wilson loop correlators in the two models respectively and observe that their expectation values follow the area law and show string breaking behavior. Further we calculate the catenary solution for the two cases and try to find indications for minimal surface behavior or string surface tension leading to string constriction.%However, even though we are using exponential error reduction our data does not allow to reveal the latter effects.
\end{abstract}

\pacs{11.15.Ha, 12.38.Gc}

\keywords{Center Vortices, Lattice Gauge Field Theory}

\maketitle

\tableofcontents

\newpage

\section{Introduction}

Many physical principles can be formulated in terms of variational problems.
For example the least-action principle is an assertion about the nature of
motion that provides an alternative approach to mechanics completely independent
of Newton's laws. Not only does the least-action principle offer a means of
formulating classical mechanics that is more flexible and powerful than
Newtonian mechanics, but also variations on the least-action principle have
proved useful in general relativity theory, quantum field theory, and particle
physics. As a result, this principle lies at the core of much of contemporary
theoretical physics.

Calculus of variations seeks to find the path, curve, surface, etc., for which a given function has a stationary value, which, in physical problems, is usually a minimum or maximum. In this article we want to use calculus of variations to analyze the (minimal) area law behavior of quark confinement in meson-meson correlators. For this behavior the non-Abelian behavior of the gauge group is essential, and especially the center vortex model of confinement has proved to be very successful in predicting the area law. 

Center vortices~\cite{'tHooft:1977hy,Vinciarelli:1978kp,
Yoneya:1978dt,Cornwall:1979hz,Mack:1978rq,Nielsen:1979xu} are topological
defects associated with the elementary center degree of freedom of the QCD gauge
field. In $D$-dimensional space-time, they are (thickened) ($D-2$)-dimensional
chromo-magnetic flux degrees of freedom and have a clear theoretical link to
confinement~\cite{'tHooft:1977hy,'tHooft:1979uj,Greensite:2003bk,Engelhardt:2004pf}.
This was confirmed by a multitude of numerical simulations, in lattice Yang-Mills theory, 
see {\it e.g.}~\cite{DelDebbio:1996mh,Langfeld:1997jx,DelDebbio:1997ke,DelDebbio:1998uu,
Kovacs:1998xm,Alexandrou:1999iy,Engelhardt:1999fd,Bertle:2001xd, Bertle:2002mm}, 
and within the infrared effective models of center vortices under investigation~
\cite{Engelhardt:1999wr,Engelhardt:2003wm,Altarawneh:2014aa,Hollwieser:2014lxa}.
Recent results~\cite{Greensite:2014gra} have also suggested that the center
vortex model of confinement is more consistent with lattice results than other
currently available models. Lattice QCD simulations indicate further that
vortices are responsible for the spontaneous breaking of chiral symmetry
($\chi$SB), dynamical mass generation and the axial $U_A(1)$ anomaly~\cite{deForcrand:1999ms,Alexandrou:1999vx,Reinhardt:2000ck2,Engelhardt:2002qs,Leinweber:2006zq,Bornyakov:2007fz,Jordan:2007ff,Hollwieser:2008tq,Hollwieser:2009wka,Bowman:2010zr,Hollwieser:2010mj,Engelhardt:2010ft,Hollwieser:2011uj,Hollwieser:2012kb,Schweigler:2012ae,Hollwieser:2013xja,Trewartha:2014ona,Brambilla:2014jmp,Nejad:2015aia,Trewartha:2015nna}, and thus 
successfully explain the non-perturbative phenomena which characterize the 
infrared sector of strong interaction physics.

In the present work we measure meson-meson correlators in effective random
center vortex models, which will be introduced in the next section. Meson and
baryon (Polyakov line) correlators were analyzed in the random vortex
world-surface model in~\cite{Altarawneh:2015bya}, showing that the correlators
follow an area law behavior. Here we investigate quadratic and circular Wilson
(not Polyakov) loop correlators, focusing on the minimal area law. Motivated by
the minimal surface of revolution problem we want to analyze possible signs of
catenary solutions.

The hanging chain or catenary problem (the world “catenary” comes from the Latin
word “catena” meaning chain) was first posed in the Acta Eruditorium in May 1690
by Jacob Bernoulli as follows: ”To find the curve assumed by a loose string hung
freely from two fixed points”. Earlier, Galileo
mistakenly conjectured that the curve was a parabola. Later Joachim Jung proved
that the curve cannot be a parabola but without presenting any solution of the
real curve. In June 1691 there were three solutions published, from Leibniz,
Huygens and Johann Bernoulli brother of Jacob. Even though these mathematicians
approached this problem in three different ways they concluded that the curve
was the hyperbolic cosine, which then came to be known as the catenary. 

The catenoid is a surface of revolution with minimum area: Any surface of
revolution generated by a different curve joining the same endpoints and having
the same length, must have a larger surface area. The proof of this fact is less
elementary and involves calculus of variations (see e.g.~\cite{Bliss:1946}). In
section~\ref{sec:cat} we derive the so-called catenary solution using the
calculus of variations, for both, standard circular and quadratic Wilson loops
and compare the results with numerical measurements in section~\ref{sec:mes}.
We finish with concluding remarks in~\ref{sec:con}.

\section{Random center vortex models}\label{sec:model}

Center vortices are expected to behave as random lines (for $D = 3$) or random
surfaces (for $D = 4$). The magnetic flux carried by the vortices is quantized
in units which are singled out by the topology of the gauge group, such that the
flux is stable against small local fluctuations of the gauge fields. The vortex
model of confinement states that the deconfinement transition is simply a
percolation transition of these chromo-magnetic flux degrees of freedom. The
following infrared effective models in $D=3$ and $D=4$ dimensions have confirmed
these expectations and in the latter case also account for topological
susceptibility and the (quenched) chiral condensate.

\subsection{Random vortex world-line model}\label{sec:model1}

In~\cite{Altarawneh:2014aa,Hollwieser:2014lxa} we introduced a model of random
flux lines in $D=2+1$ space-time dimensions. The lines are composed of straight
segments connecting nodes randomly distributed in continuous three-dimensional
space. Each node is connected to two lines, hence the configuration only
consists of closed vortex clusters. The physical space in which the vortex lines
are defined is a cuboid $L_S^2\times L_T$ with "spatial" extent $L_S$,
"temporal" extent $L_T$ and periodic boundary conditions in all directions.
Within this paper we use volumes with $L_S=L_T=16$, where finite size effects
are under control. The vortex length $L$ between two nodes is restricted to a
certain range $L_{min}<L<L_{max}$. This range in some sense sets the scale of
the model; for practical reasons we choose a scale of $L\approx1$, {\it i.e.}
$L_{min}=0.3$ and $L_{max}=1.7$ in these dimensionless units. All updates
resulting in vortex lengths $L$ out of the range $L_{min}<L<L_{max}$ are
rejected. 

An ensemble is generated by Monte Carlo methods, starting with a random initial
configuration.  Allowance is made for nodes moving as well as being added or
deleted from the configurations during Monte Carlo updates. A Metropolis
algorithm is applied to all updates using the action $S= \alpha L + \gamma
\varphi^2,\label{eq:act}$ with action parameters $\alpha=0.11$ and $\gamma=0.33$ for the
vortex length $L$ and the vortex angle $\varphi$ between two adjacent segments
respectively. The difference of the action of the affected nodes before and
after the update determines the probability of the update to be accepted. The
move update moves the current node by a random vector of maximal length
$r_m=4L_{min}$, it affects the action of three nodes, the node itself and its
neighbors. The add update adds a node at a random position within a radius
$r_a=3L_{min}$ around the midpoint between the current and the next node. The
action before the update is given by the sum of the action at the current and
the next node, while the action after the update is the sum of the action at the
current, the new and the next node. Deleting the current node, on the other
hand, affects three nodes before the update and only two nodes after the update.
Therefore the probability for the add update is in general much smaller than for
the delete update and the vortex structure would soon vanish if both updates
were tried equally often. Hence, the update strategy is randomized to move a
node in two out of three cases ($66\%$), and apply the add update about five
times more often than the delete update ($28\%$ vs. $6\%$). A density parameter
$d$ is restricting the number of nodes in a certain volume. The add update is
rejected if the number of nodes within a $3\times3\times3$ volume around the new
node exceeds the density parameter $d$. 

Furthermore, Monte Carlo updates disconnecting and fusing vortex lines were
implemented, {\it i.e.}, when two vortices approach each other, they can
reconnect or separate at a bottleneck. The ensemble therefore will contain not a
fixed, but a variable number of closed vortex lines or "vortex clusters".  Given
that the deconfining phase transition is a percolation transition, such
processes play a crucial role in the vortex picture. If the current node is not
deleted, all nodes around the current node are considered for reconnections. The
reconnection update causes the cancellation of two close, nearly parallel vortex
lines and reconnection of the involved nodes with new vortex lines. The
reconnection update is also subjected to the Metropolis algorithm, considering
the action of the four nodes involved. 

The model was shown to exhibit both a low-temperature confining phase and a
high-temperature deconfined phase, as well as phase transitions with respect to
vortex density and segment length $L$. The predictions of the model for the
spatial string tension in the deconfined phase quantitatively match
corresponding $SU(2)$ lattice Yang-Mills results.

\subsection{Random vortex world-surface model}\label{sec:model2}

The random vortex world-surface model was introduced in~\cite{Engelhardt:1999wr,
Engelhardt:2000wc} based on the notion that the Yang-Mills vacuum is populated
by collective magnetic vortex degrees of freedom which are represented by closed
two-dimensional world-surfaces in four-dimensional (Euclidean) space-time. The
chromomagnetic flux carried by the vortices is quantized according to the center
of the gauge group. The vortex world-surfaces are treated as random surfaces, an
ensemble of which in practice is generated using Monte Carlo methods on a
hypercubic lattice; the surfaces are composed of elementary squares (plaquettes)
on that lattice. The spacing of the lattice is a fixed physical quantity,
related to an intrinsic thickness of the vortex fluxes, and represents the
ultraviolet cutoff inherent in any infrared effective framework. The action
governing the ensemble is related to the surface curvature: If two elementary
squares which are part of a vortex surface share a lattice link but do not lie
in the same plane, this costs an action increment $c$, where the parameter is
set to $c=0.24$, such as to reproduce the $SU(2)$ Yang-Mills ratio of the
deconfinement temperature to the square root of the zero-temperature string
tension, $T_C/\sqrt{\sigma_0}=0.69$~\cite{Engelhardt:1999wr}. 

Physically, the random vortex surfaces represent quantized chromomagnetic flux.
This means that they contribute in a characteristic way to Wilson loops; if one
chooses an area spanning a given Wilson loop - the choice of area is immaterial
due to the continuity of flux - then for each time a vortex world-surface
pierces that area, the Wilson loop acquires a phase factor corresponding to the
center of the gauge group. For the case of $SU(2)$ color treated in this paper,
the Wilson loop picks up a $(-1)=\exp(\pm i\pi)$. Note that Wilson loops in the
4D model are defined on a lattice dual to the one on which vortices are defined,
i.e. on a lattice shifted by the vector $(a/2, a/2, a/2, a/2)$, where $a$
denotes the lattice spacing. Thus, the notion of a vortex piercing a Wilson loop
area is unambiguous. 

The vortex ensemble is generated subject to the constraint of continuity of flux
(modulo $2\pi $, {\it i.e.}, modulo Dirac
strings~\cite{Engelhardt:2000wc,Engelhardt:2002qs,Engelhardt:2003wm}), forcing
thee vortex surfaces to be closed, as already mentioned above. As described in
detail in~\cite{Engelhardt:2003wm}, continuity of flux in practice is guaranteed
during the generation of the vortex world-surface ensemble by performing updates
simultaneously on the six squares making up the surface of an elementary
three-dimensional cube in the lattice. This is done in a way which corresponds
to superimposing the (continuous) flux of a vortex of the shape of the
elementary cube surface onto the flux previously present.

For the $SU(2)$ gauge group, the model was shown to exhibit both a
low-temperature confining phase as well as a high-temperature deconfined
phase~\cite{Engelhardt:1999wr}, separated by a second-order phase
transition~\cite{Engelhardt:2003wm}. The predictions of the model for the
spatial string tension in the deconfined phase~\cite{Engelhardt:1999wr},  the
topological susceptibility~\cite{Engelhardt:2000wc} and the (quenched) chiral
condensate~\cite{Engelhardt:2002qs} quantitatively match corresponding $SU(2)$
lattice Yang-Mills results. Building on this initial progress, the random vortex
world-surface model was extended to the case of $SU(3)$ color. Studies of the
confinement properties yielded a weakly first-order deconfinement phase
transition~\cite{Engelhardt:2003wm}, also seen in the vortex free
energy~\cite{Quandt:2004gy}, which represents an alternative order parameter for
confinement. Furthermore, a $Y$ law for the baryonic static potential was
observed~\cite{Engelhardt:2004qq} and topological susceptibility was studied
in~\cite{Engelhardt:2010ft}, which is instrumental in determining, via the
$U_A(1)$ anomaly, the mass of the $\eta'$ meson. Spurred by investigations of
Yang-Mills theories with a wider variety of gauge groups, aiming at a better
understanding of possible confinement mechanisms~\cite{Holland:2003kg}, the
confinement properties of the random vortex world-surface model were
subsequently also studied for $SU(4)$ color~\cite{Engelhardt:2005qu} and $Sp(2)$
color~\cite{Engelhardt:2006ep}. These studies showed that the vortex picture can
accomodate such diverse color symmetries, while indicating the limitations of
the very simple effective dynamics which had proven adequate in the $SU(2)$ and
$SU(3)$ cases. Finally, as already mentioned above, meson and baryon correlators
were analyzed in the random vortex world-surface model for the latter cases 
in~\cite{Altarawneh:2015bya}, showing that the correlators follow an area law
behavior and are in qualitative agreement with lattice studies.

\section{The Catenary Solution}\label{sec:cat}

\begin{figure}[h]
	\centering
	a)\includegraphics[width=.44\linewidth]{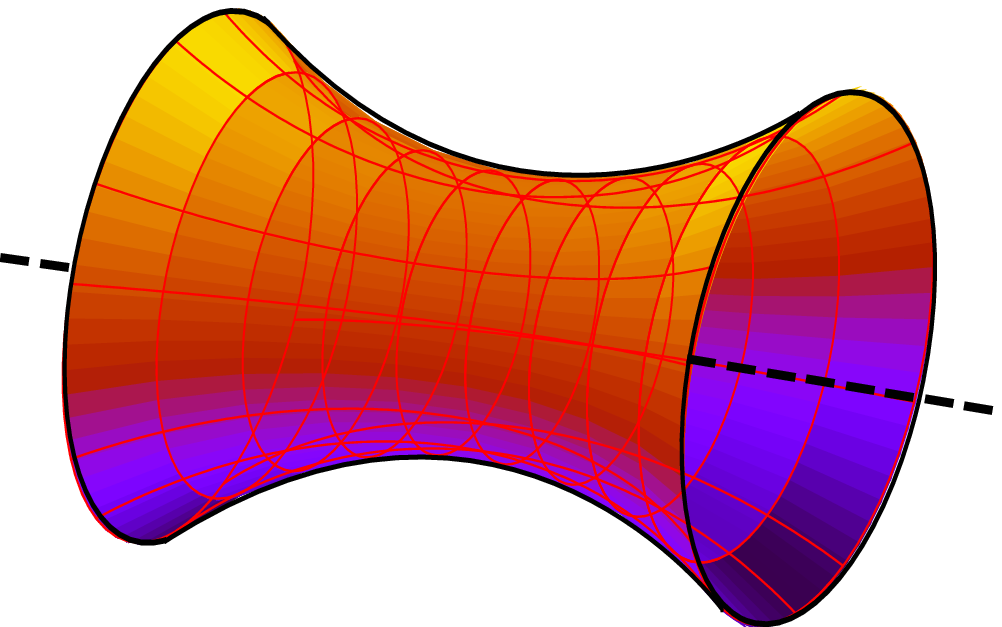}$\qquad$$\qquad$
	b)\includegraphics[width=.38\linewidth]{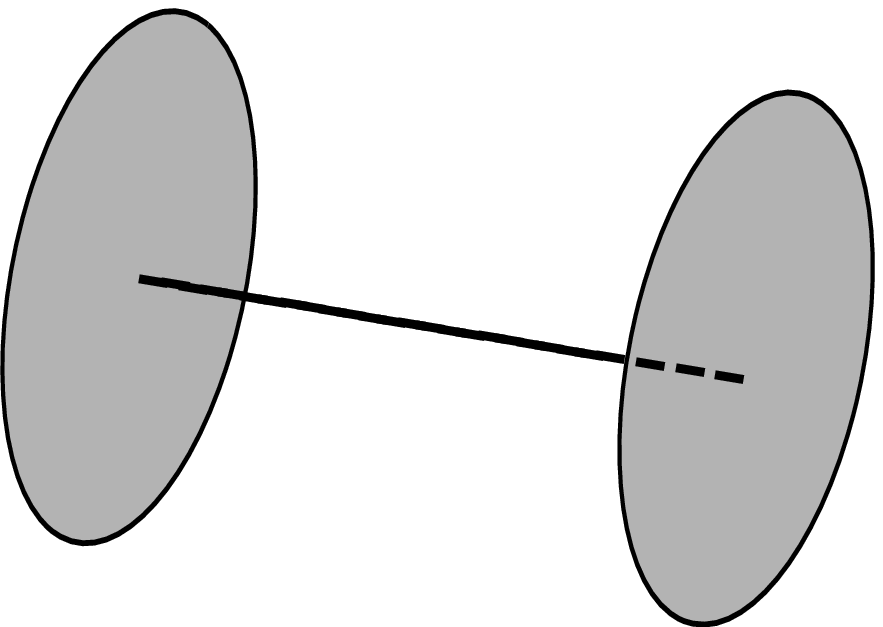}
	\caption{Solutions of the catenary problem for circular Wilson loops:
	a) Catenoid and b) Goldschmidt's discontinuous solution.}
	\label{fig:cat}
\end{figure}

A power line hanging between two poles shows us a curve called the catenary. 
The shape of this natural curve can be derived from a differential equation 
describing the physics behind a uniform flexible chain hanging by its own weight. 
The problem was also proposed by W. Symmond in~\cite{Symmond:1894} in 1894. J. C.
Nagel gave a solution in~\cite{Nagel:1895} that is easy to describe if we accept
the reasonable fact that the catenary is the curve with the lowest center of
gravity. The problem can be formulated in the same way as minimizing the surface
of revolution of a curve between two fixed points, well known from calculus of
variation, see {\it e.g.}~\cite{wolfram}: 

Given $P_1$, $P_2$ in the plane, find a curve $y(x)$ from $P_1$ to $P_2$ such
that the surface of revolution obtained by revolving the curve about the
$x$-axis has minimum surface area. In other words, minimize
$2\pi\int_{y(x_1)}^{y(x_2)}yds$ with $y(x_i)=P_i$ and $ds$ the line segment
along the curve $y(x)$. If $P_1$ and $P_2$ are not too far apart, relative to
$x_2-x_1$ then the solution is a Catenary (the resulting surface is called a
Catenoid), see Fig.~\ref{fig:cat}.

Otherwise the solution is Goldschmidt's discontinuous solution (discovered in
1831) obtained by revolving the curve which is the union of three lines: the
vertical line from $P_1$ to the point $(x_1,0)$, the vertical line from $P_2$ to
$(x_2,0)$ and the segment of the $x$-axis from $(x_1,0)$ to $(x_2,0)$. This
example also straightforwardly translates to the case of the minimal area law
for the Wilson loops, {\it i.e.}, if the two Wilson loops are far apart, we only
measure the product of their expectation values or the string tensions of the
two individual meson pairs. However, if the meson pairs approach each other, we
measure mixed states all four mesons interact with each other and at some point
the strings will flip and form new bound states. The question now is if we can
interpret these mixed state measurements with catenary solutions. Therefore we
derive the equations for the set of boundary conditions of our problem, starting
with the circular Wilson loop.

\subsection{Circular Wilson loops}\label{ssec:cwl}

For circular Wilson loops the surface of revolution problem is straightforward.
We want to find a curve from a point $(-D/2,R)$ to a point $(D/2,R)$, where $R$
is the radius of the Wilson loop and $D$ is the distance between two Wilson
loops, which, when revolved around the $x$-axis yields  a surface of smallest
surface area $A$, {\it i.e.}, the minimal surface. The area element is 
\begin{equation}
dA=2\pi y ds=2\pi y\sqrt{1+y'^2}dx,
\end{equation}
so the surface area is
\begin{equation}
A=2\pi\int_{-D/2}^{D/2}y\sqrt{1+y'^2}dx=4\pi\int_0^{D/2}y\sqrt{1+y'^2}dx,\label{eq:area}
\end{equation}
and the quantity to minimize is
\begin{equation}
f=y\sqrt{1+y'^2}.\label{eq:min}
\end{equation}
This can be solved analytically by using the Beltrami identity, see
appendix~\ref{app:cat}, and gives the solution (\ref{eq:los})
\begin{equation}
y=a\cosh(\frac{x-b}{a}),\label{eq:sol}
\end{equation}
with two constants $a$ and $b$ to be determined from the boundary conditions
\begin{equation}
R=a\cosh(\frac{-D/2-b}{a})=a\cosh(\frac{D/2-b}{a}).
\end{equation}
Since $\cosh(-x)=\cosh(x)$ we get $D/2-b=D/2+b$, hence $b=0$, as it must by symmetry, and the minimal surface solution reduces to 
\begin{equation}
y=a\cosh(x/a) \quad\mbox{with the boundary condition}\quad R=a\cosh(D/2a),\label{eq:bcs}
\end{equation}
to determine the constant $a$. But for certain values $R$ and $D$, this equation
has no solution. The mathematical interpretation of this fact is that the
surface breaks and forms circular disks in each ring to minimize area, {\it
i.e.} the Goldschmidt solution. Physically it simply means that the minimal area
is given by the two Wilson loops rather than any surface of revolution, the two
interpretations however do not agree exactly. In appendix~\ref{app:cvsg} we
derive that we only obtain catenary solutions for $D/R<1.325$
(\ref{eq:cmax}), but not all solutions are absolute minima of our minimal area
problem. The surface area of the minimal catenoid is given by (\ref{eq:area})
\begin{equation}
A=4\pi\int_0^{D/2}y\sqrt{1+y'^2}dx,
\end{equation}
but since according to Eqs. (\ref{eq:sol}) and (\ref{eq:4})
\begin{equation}
y=a\cosh(x/a)=\sqrt{1+y'^2}a,
\end{equation}
\begin{eqnarray}
A&=&\frac{4\pi}{a}\int_0^{D/2}y^2dx=4\pi a\int_0^{D/2}\cosh^2(x/a)dx\\
&=&\pi a^2\left[\sinh(\frac{2x}{a})+\frac{2x}{a}\right]_0^{D/2}=\pi a^2\left[\sinh(D/a)+D/a\right].\label{eq:circA}
\end{eqnarray}
The surface area of the catenoid equals that of the Goldschmidt solution when (\ref{eq:circA}) equals the area of two disks, the circular Wilson loops with radius $R$, {\it i.e.}, 
\begin{eqnarray}
\pi a^2\left[\sinh(D/a)+D/a\right]&=&2\pi R^2\\
\sinh(D/2a)\cosh(D/2a)+D/2a&=&R^2/a^2\\
\cosh(D/2a)\sqrt{\cosh^2(D/2a)-1}+D/2a&=&R^2/a^2.\label{eq:ceqg}
\end{eqnarray}
Defining $u\equiv R/a=\cosh(D/2a)$ from (\ref{eq:bcs}) and plugging it in (\ref{eq:ceqg}) gives $u\sqrt{u^2-1}+\cosh^{-1}u=u^2$ which has a solution $u\approx1.211$. The value of $D/R$ for which $A_{catenary}=A_{2disks}$ is therefore $D/R=2\cosh^{-1}u/u\approx1.055$. For $D/R\in(1.055,1.325)$, the catenary solution has larger area than the two Wilson loops, so it exists only as a local minimum. For $D/R<1.055$ we evaluate the area (\ref{eq:circA}) numerically with $a$ from the boundary condition (\ref{eq:bcs}).

\subsection{Quadratic Wilson loops}

\begin{figure}[h]
	\centering
	a)\includegraphics[width=.4\linewidth]{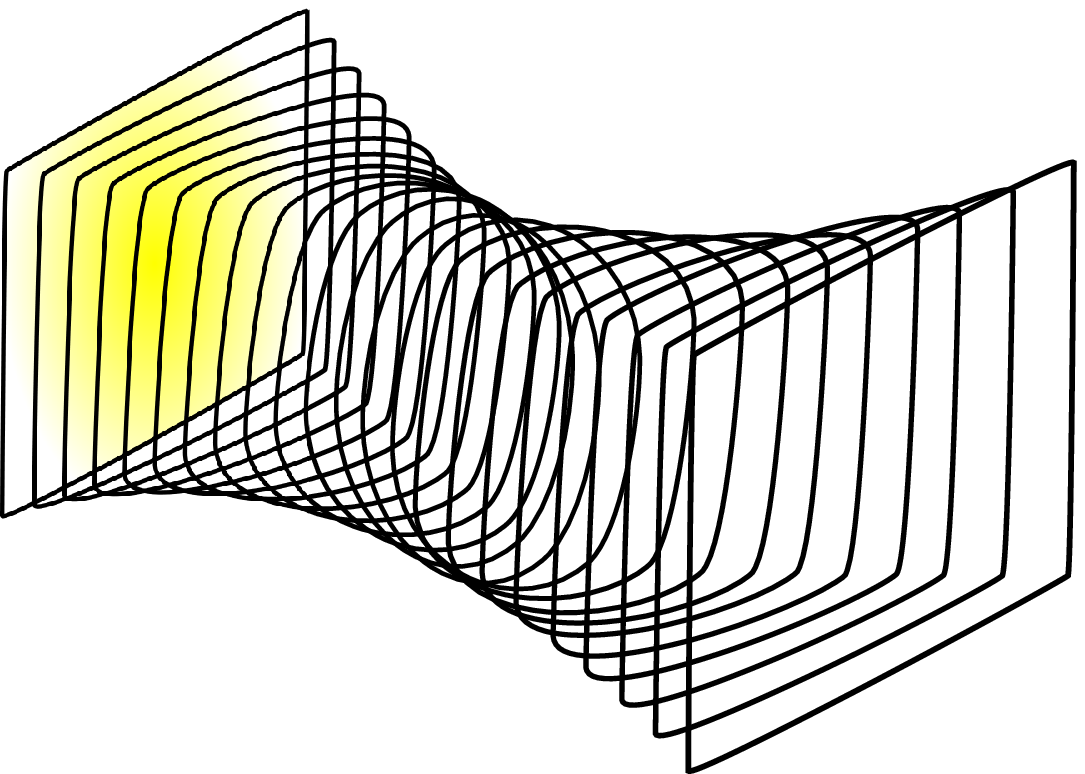}$\qquad$$\qquad$
	b)\includegraphics[width=.4\linewidth]{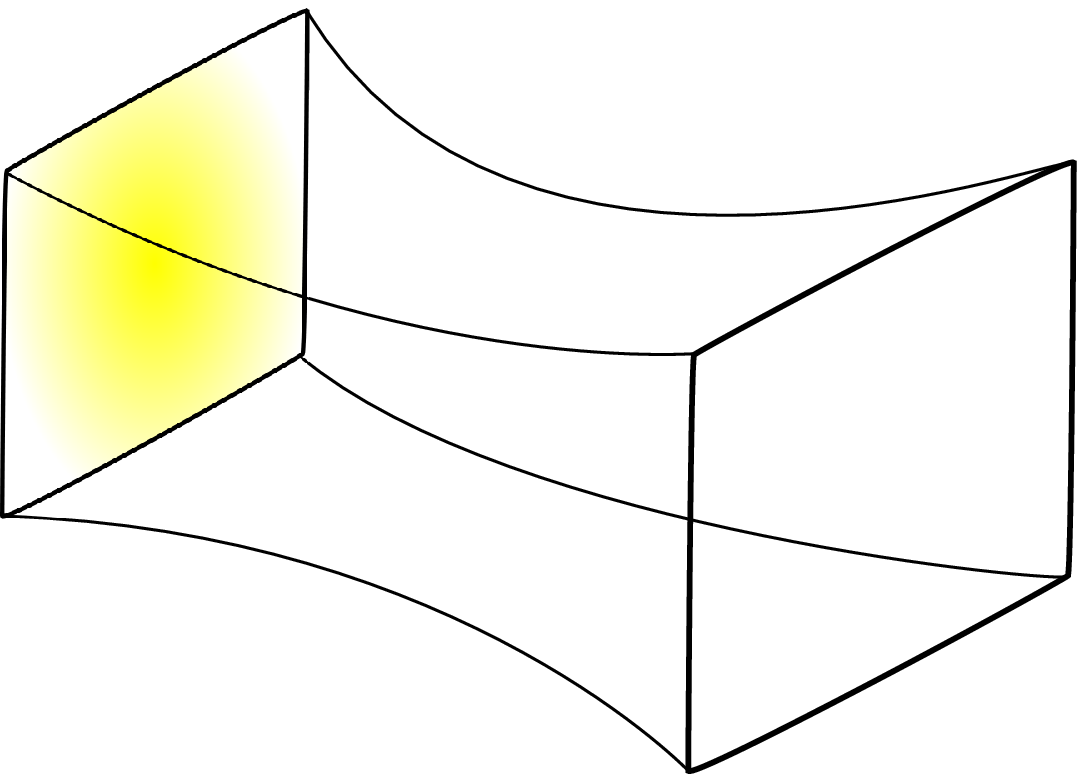}
	\caption{Minimal surface solutions for rectangular Wilson loops.}
	\label{fig:quat}
\end{figure}

For quadratic Wilson loops the minimal surface solution would probably look like Fig.~\ref{fig:quat}a, {\it i.e.}, an interpolation of Fig.~\ref{fig:quat}b and the catenary solution in Fig.~\ref{fig:cat}a. We are going to approximate the problem however with the minimal area of Fig.~\ref{fig:quat}b and therefore have to minimize the four lateral surfaces.

This can be done by identifying each of the four lateral surfaces with a
catenoid, {\it i.e.}, we unroll the circular solution in Fig.~\ref{fig:cat}a and
glue four of them together to obtain the minimal surface solution for
rectangular Wilson loops shown in Fig.~\ref{fig:quat}b. Starting from the
solution (\ref{eq:los}) derived in appendix~\ref{app:cat}, we can set the
integration constant $b=0$ due to symmetry as above, to obtain
\begin{equation}
y=a\cosh(x/a) \quad\mbox{and}\quad R/2\pi=a\cosh(D/2a),\label{eq:bqs}
\end{equation}
where the boundary condition is given by the fact that the perimeter of the
catenoid circle at $\pm D/2$ (after unrolling) has to be equal to the side length
$R$ of the quadratic Wilson loop. This changes the range of possible catenary
solutions and yields an upper bound $D/R<0.88$ (\ref{eq:qmax}). The total surface is now four
times the catenoid (\ref{eq:circA}), {\it i.e.}
\begin{equation}
A=4\pi a^2\left[\sinh(D/a)+D/a\right],\label{eq:quatA}
\end{equation}
which equals the Wilson loop areas for $D/R=0.59$, hence we again evaluate the
minimal area below this value numerically with $a$ from the boundary condition
(\ref{eq:bqs}) and compare it to the measurements in the next section.

\section{Measurements \& Results}\label{sec:mes}

We measure meson-meson potentials in the background of random $Z(2)$ center vortex
lines or surfaces in $D=3$ and $D=4$ space-time dimensions. In particular, we
measure the correlator of two (flat) Wilson loops with circular and quadratic
shape, respectively. In fact, these Wilson loops, especially the circular shape,
may not represent the perfect observable for measuring meson-meson potentials,
as creation and annihilation processes of the quark--anti-quark pairs can not be
neglected. But our main interest lies in the comparison with catenary solutions
and therefore these shapes are the ideal candidates.

\subsection{Quadratic Wilson loop correlators in the 4D vortex surface model}

We measure the correlator of two (flat) Wilson loops of size $R\times R$ at a
distance $D$ on $16^4$ lattices of random center vortex world-surfaces. The
$Z(2)$ vortices contribute a center element $(-1)=\exp(\pm i\pi)$ when they
pierce a Wilson loop. The piercing is unambiguous as the vortices live on the
dual lattice, as discussed in section~\ref{sec:model2}. The lattice string
tension of the model is $\sigma a^2=0.755$~\cite{Engelhardt:1999wr} gives the
scale between (minimal) area and the observable. To reduce the numerical noise
contaminating the measurements as far as possible, the exponential noise
reduction technique introduced by L\"uscher and Weisz~\cite{Luscher:2001up} was
employed as introduced in~\cite{Engelhardt:2004qq} for the random center vortex
world-surface model. It is a multilevel scheme that exploits the locality of the
theory by averaging over sub-ensembles, which can be applied to the random
vortex world-surface model as its action (and the Wilson loop) can be decomposed
into sub-lattices. In practice, we use the same number of configurations for the
individual averaging steps as detailed in~\cite{Altarawneh:2015bya} to achieve
the maximum level of accuracy. However, the correlator runs into the double precision
regime and breaks down at $-log\langle W(R,R)_xW(R,R)_{x+D}\rangle\approx24$,
{\it i.e.}, for individual Wilson loops $\langle W(R,R)\rangle\approx10^{-12}$. 

\begin{figure}[h]
	\centering
	a)\includegraphics[width=.48\linewidth]{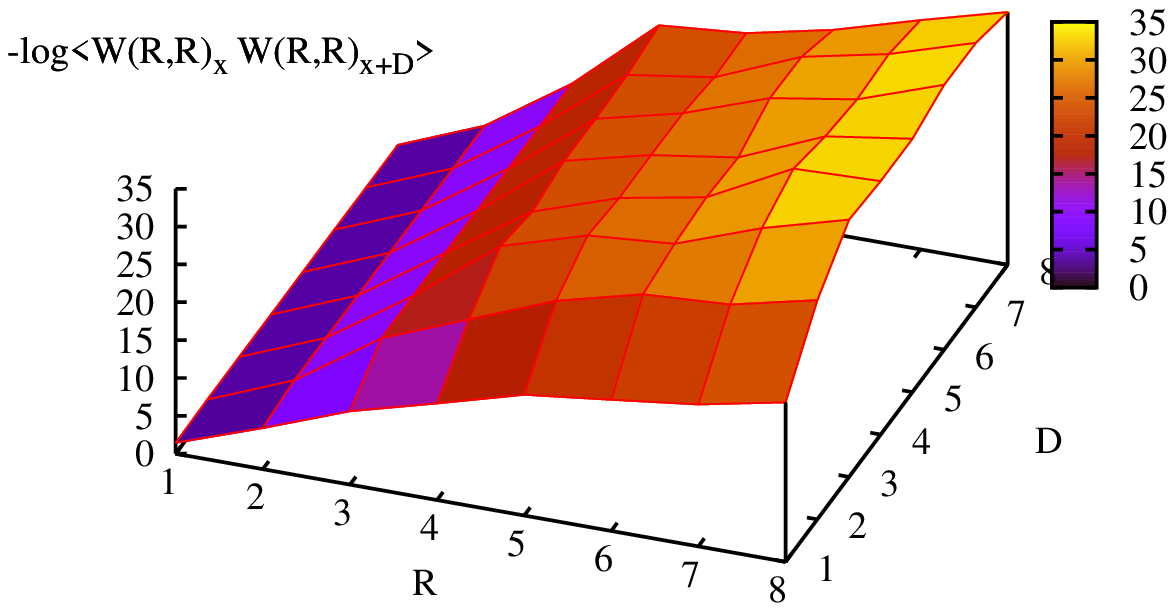}
	b)\includegraphics[width=.48\linewidth]{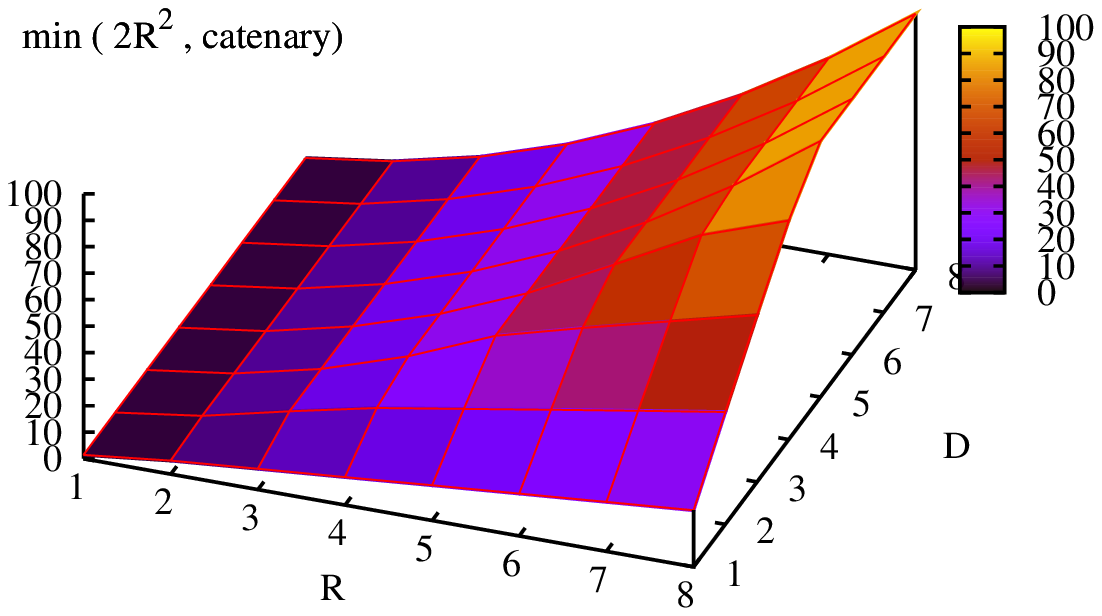}
	\caption{a) $Z(2)$ meson (quadratic Wilson loop) correlators $-log\langle
	W(R,R)_xW(R,R)_{x+D}\rangle$ for various Wilson loop sizes $R$, or distance
	between quark and anti-quark, and distances $D$ between the meson and
	anti-meson on $16^4$ lattices. b) Surface plot of the minimum of the two
	Wilson loop areas $2R^2$ and the catenoid (\ref{eq:quatA}).}
	\label{fig:3dq}
\end{figure}

We show 3D plots of the measured correlators and the minimal areas determined
from the catenary solution (\ref{eq:quatA}) or Wilson loop sizes ($2R^2$) in
Fig.~\ref{fig:3dq} and 2D cuts of the plots vs. $D$ or $R$ in
Fig.~\ref{fig:2dq}.
%and Fig.~\ref{fig:2rq} respectively. 
The data shows perfect area law behavior and we may see agreement with the
catenary solution where the coarse lattice data allows the resolution of the
effect (Figs.~\ref{fig:2dq}b and d at $D=1$ and Fig.~\ref{fig:2dq}e at $D=1$ and
$2$ resp.) and correspondingly in Figs.~\ref{fig:2dq}b and d before the signal
is lost in machine precision noise for $R$ and $D>4$.

\subsection{Circular Wilson loop correlators in the 3D vortex line model}

\begin{figure}[h]
	\centering
	a)\includegraphics[width=.45\linewidth]{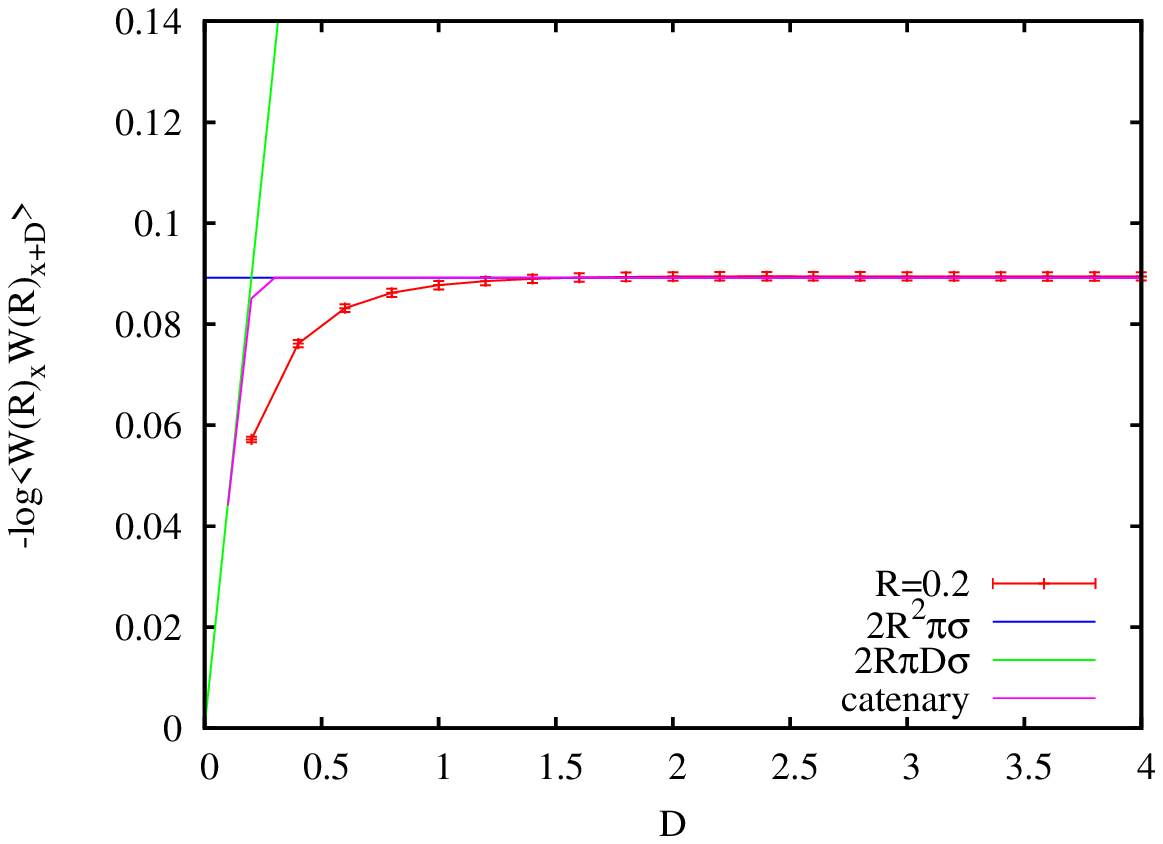}
	b)\includegraphics[width=.45\linewidth]{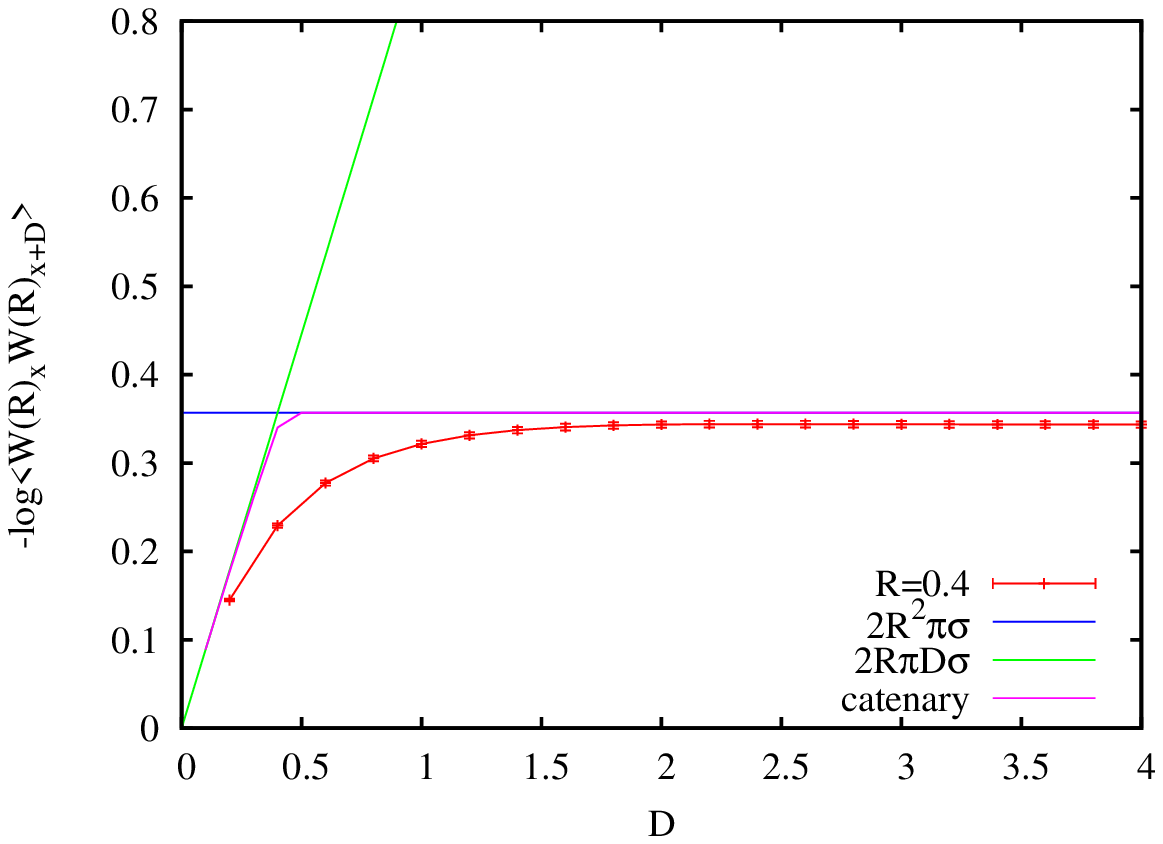}\\
	c)\includegraphics[width=.45\linewidth]{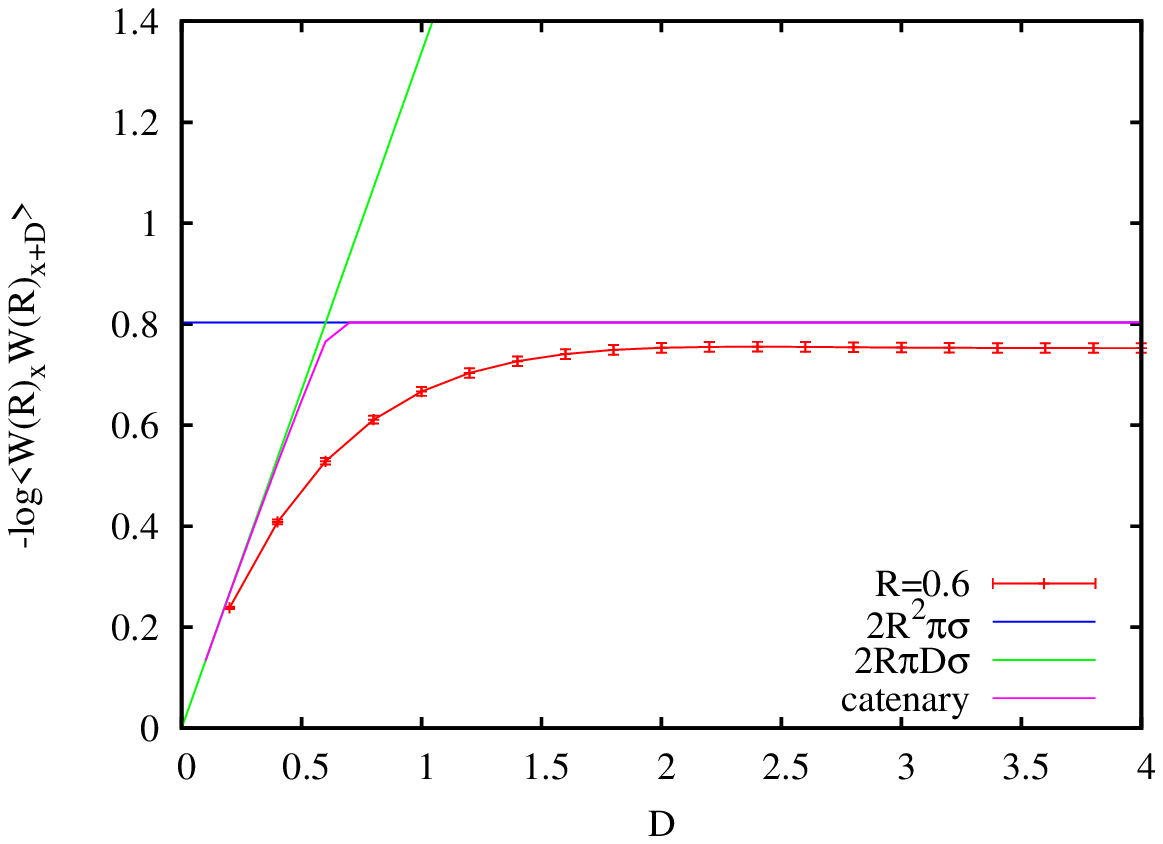}
	d)\includegraphics[width=.45\linewidth]{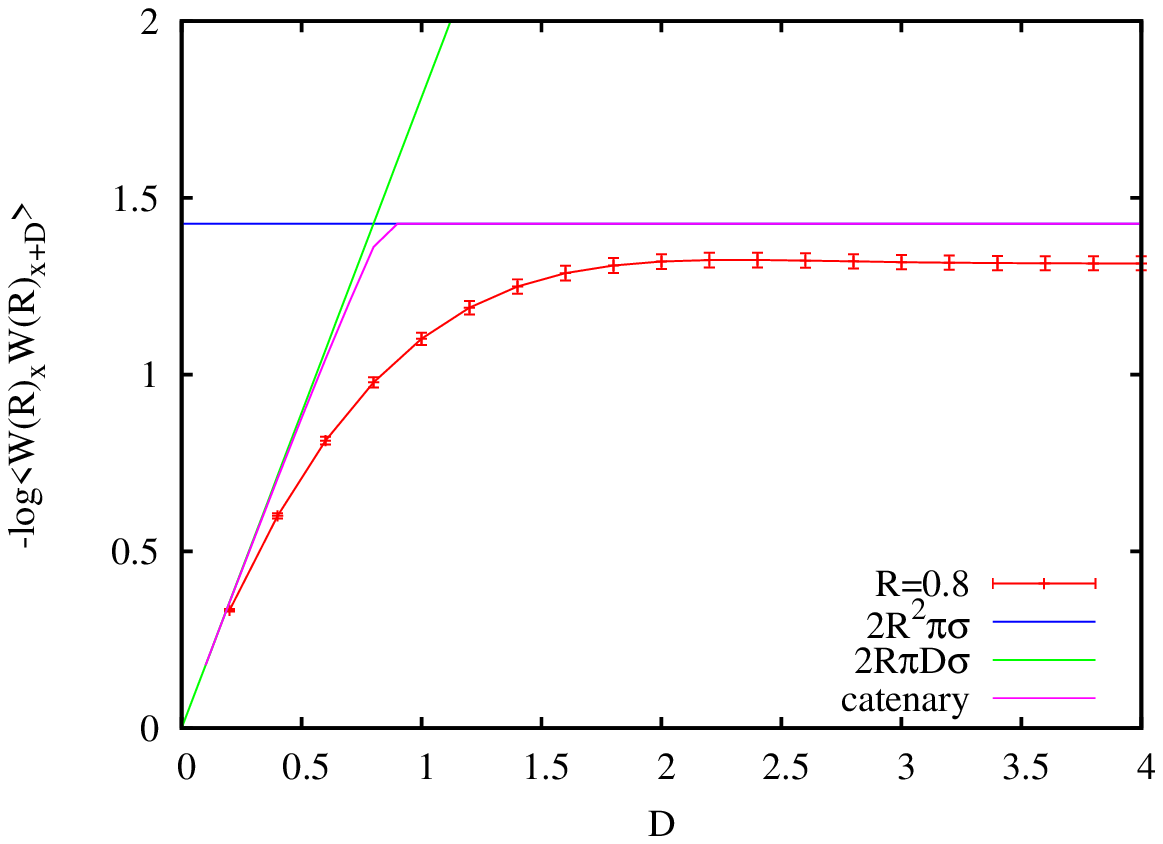}\\
	e)\includegraphics[width=.45\linewidth]{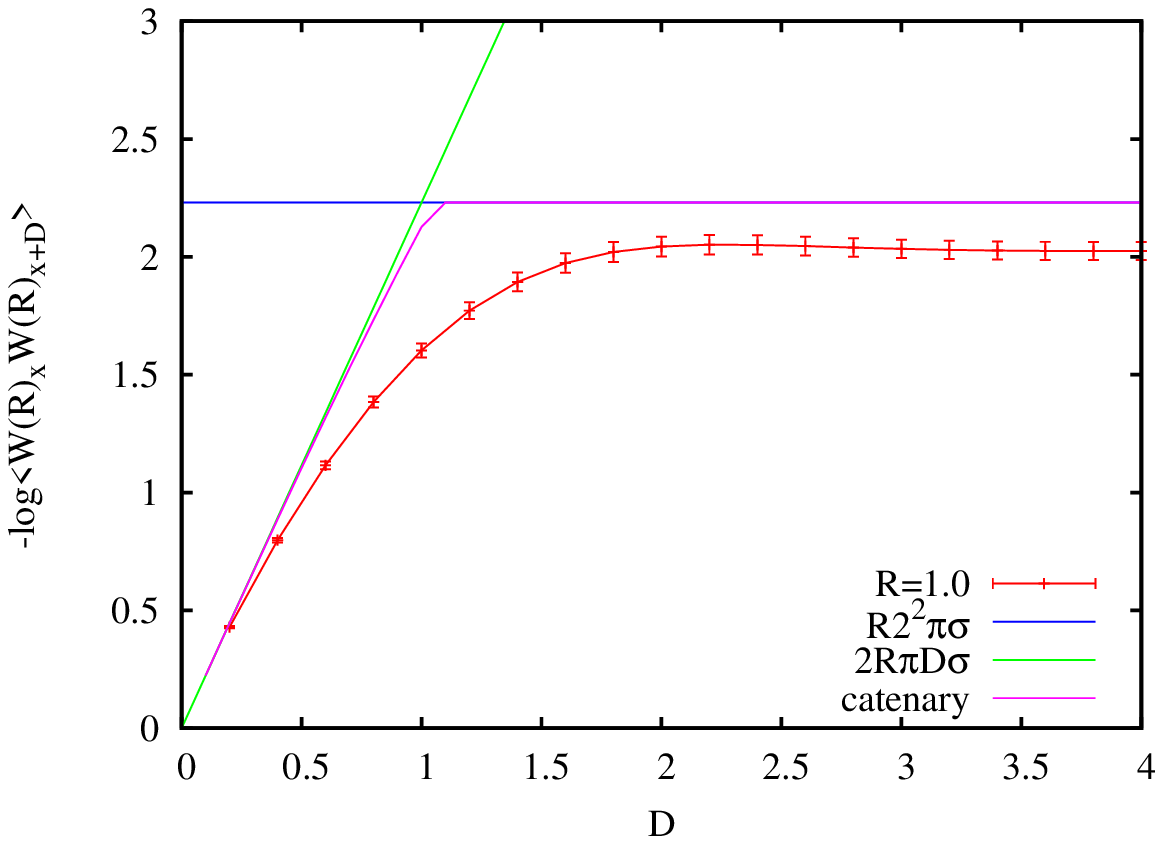}
	f)\includegraphics[width=.45\linewidth]{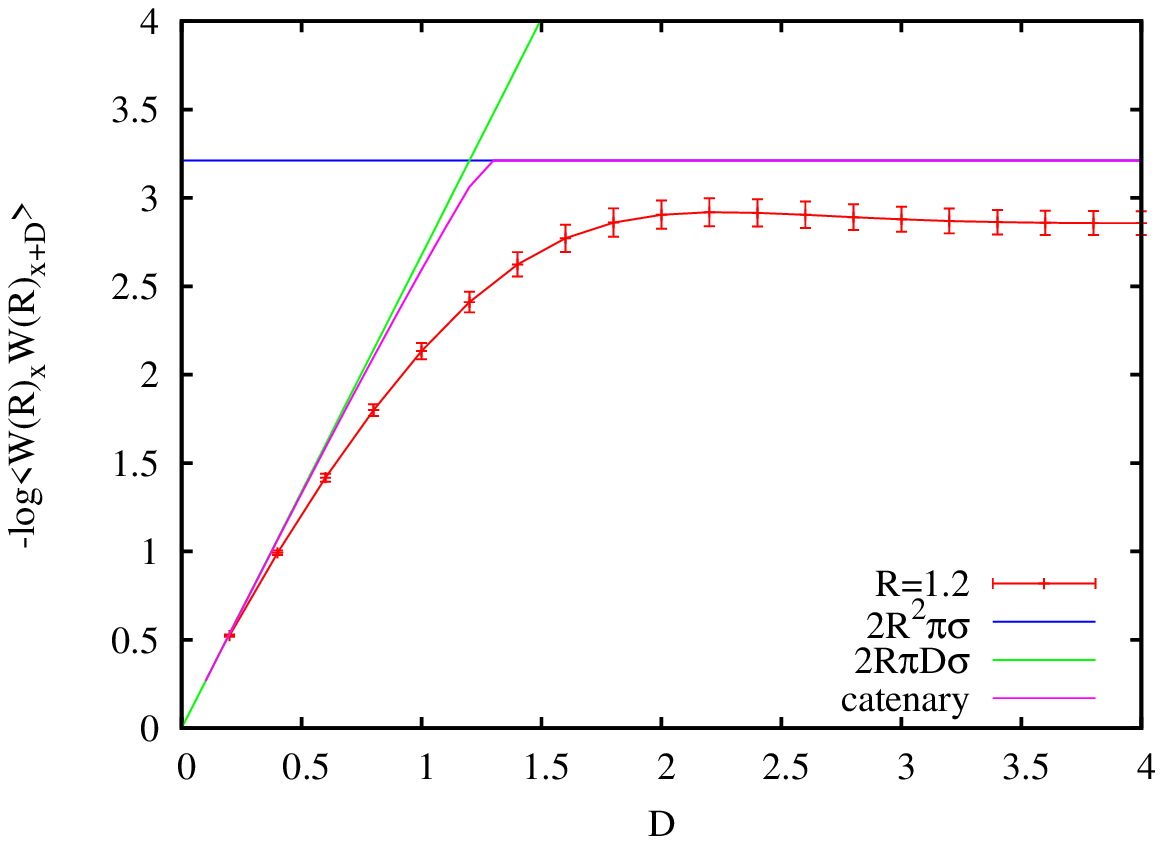}\\
	g)\includegraphics[width=.45\linewidth]{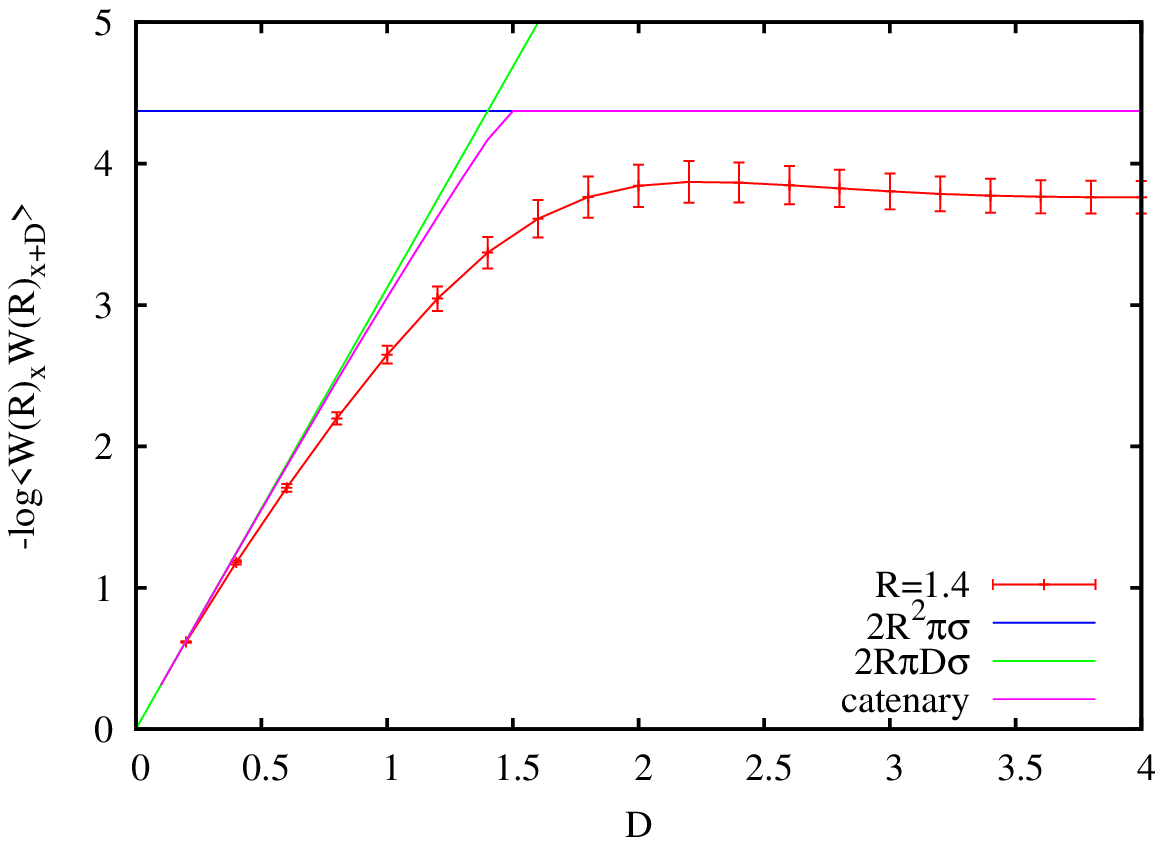}
	h)\includegraphics[width=.45\linewidth]{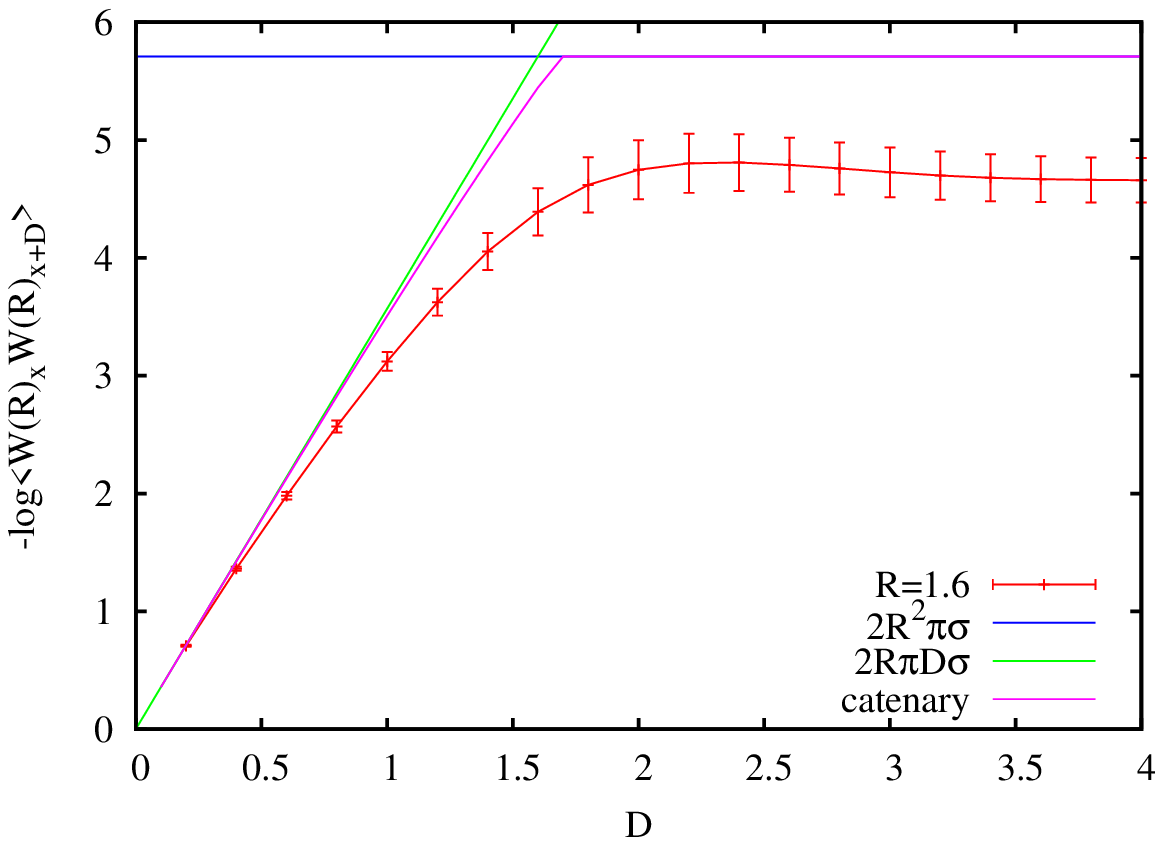}
	\caption{Circular Wilson loop correlators $-log\langle W(R)_xW(R)_{x+D}\rangle$ for various Wilson loop radii $R$ versus distance $D$ between the Wilson loops and minimal area solutions.}
	\label{fig:2rc}
\end{figure}

\begin{figure}[h]
	\centering
	a)\includegraphics[width=.45\linewidth]{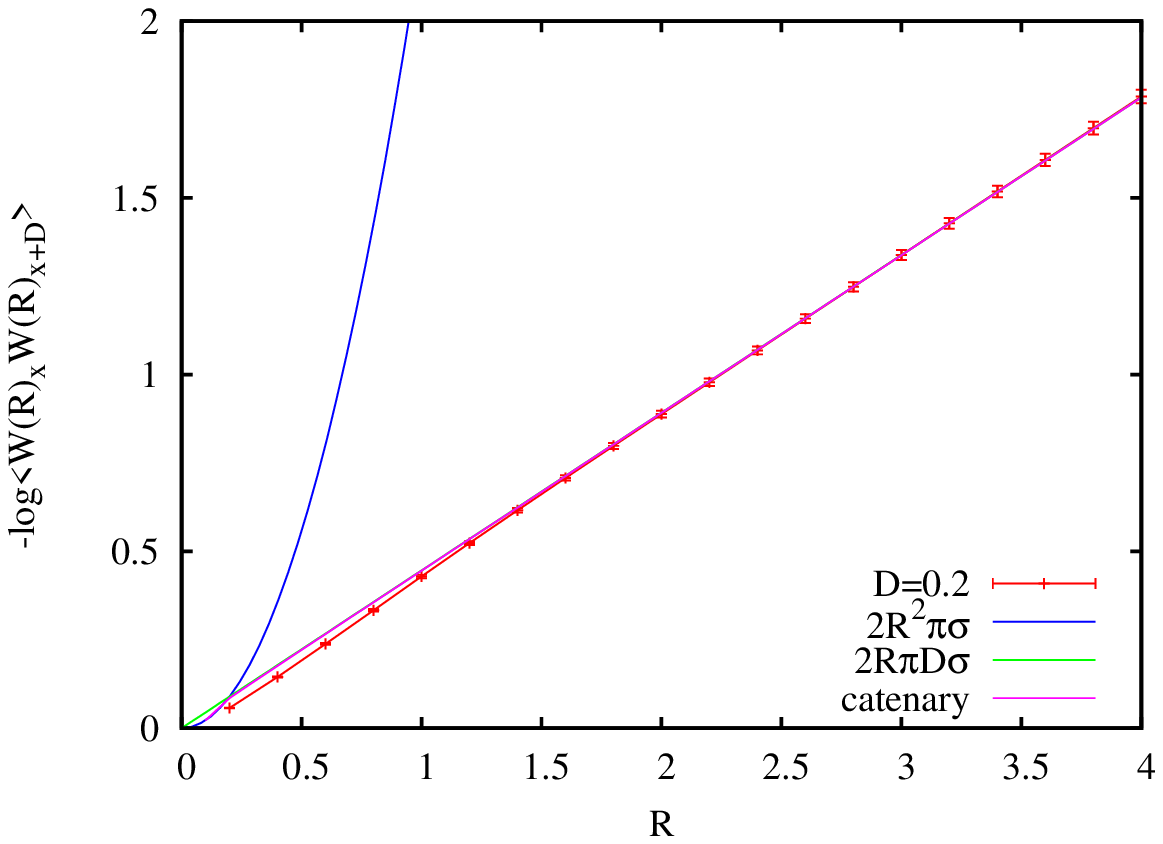}
	b)\includegraphics[width=.45\linewidth]{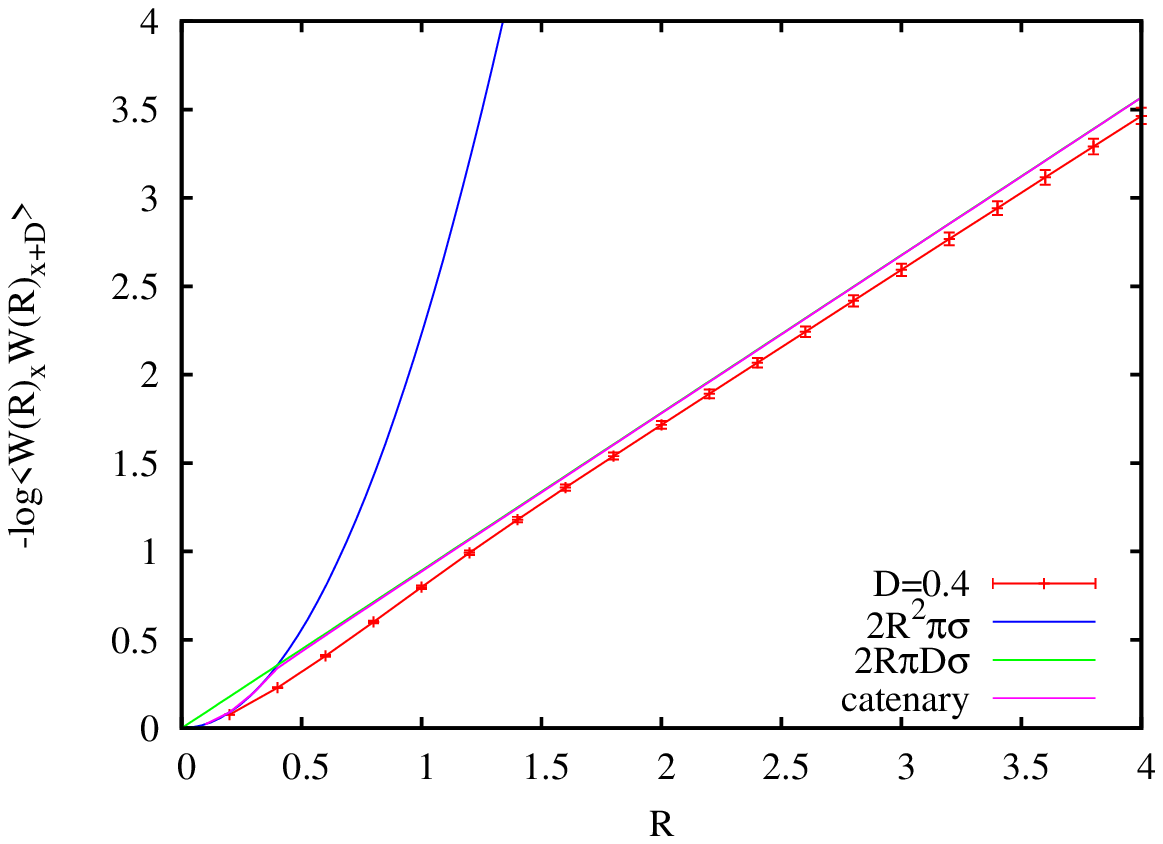}\\
	c)\includegraphics[width=.45\linewidth]{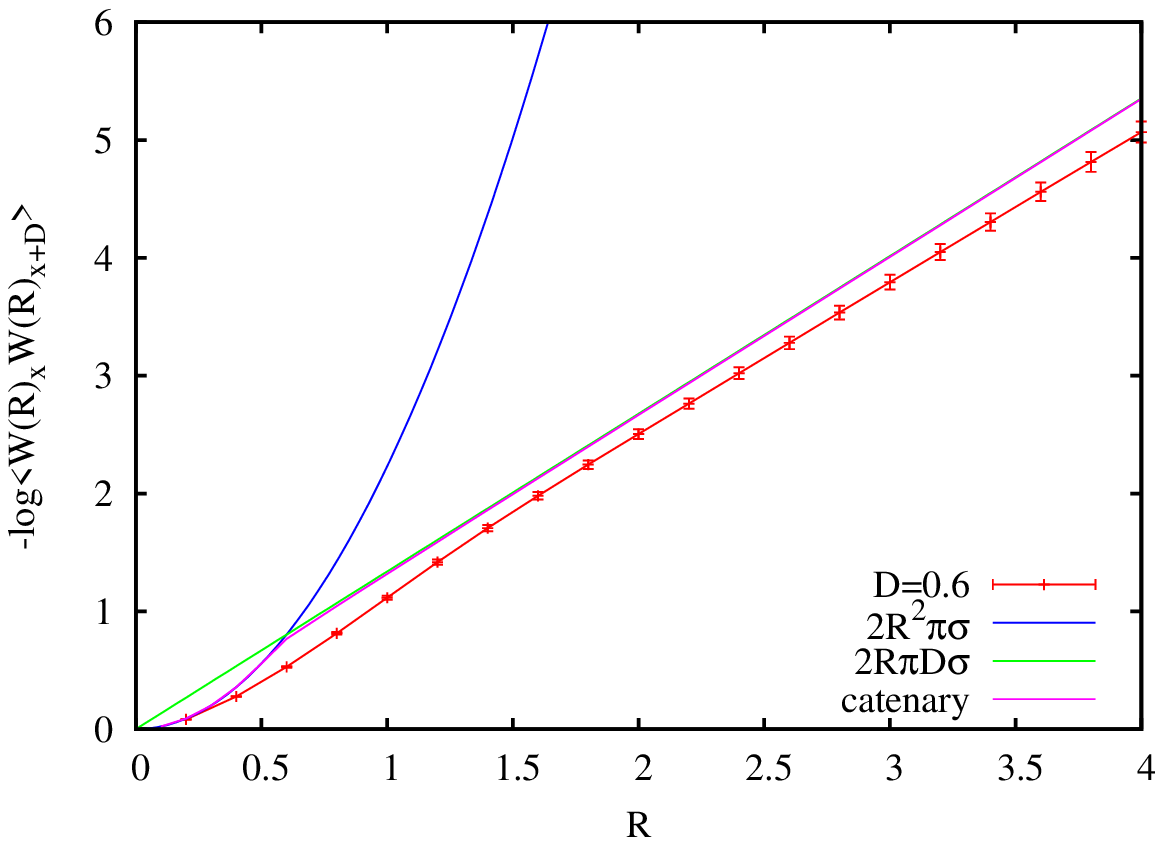}
	d)\includegraphics[width=.45\linewidth]{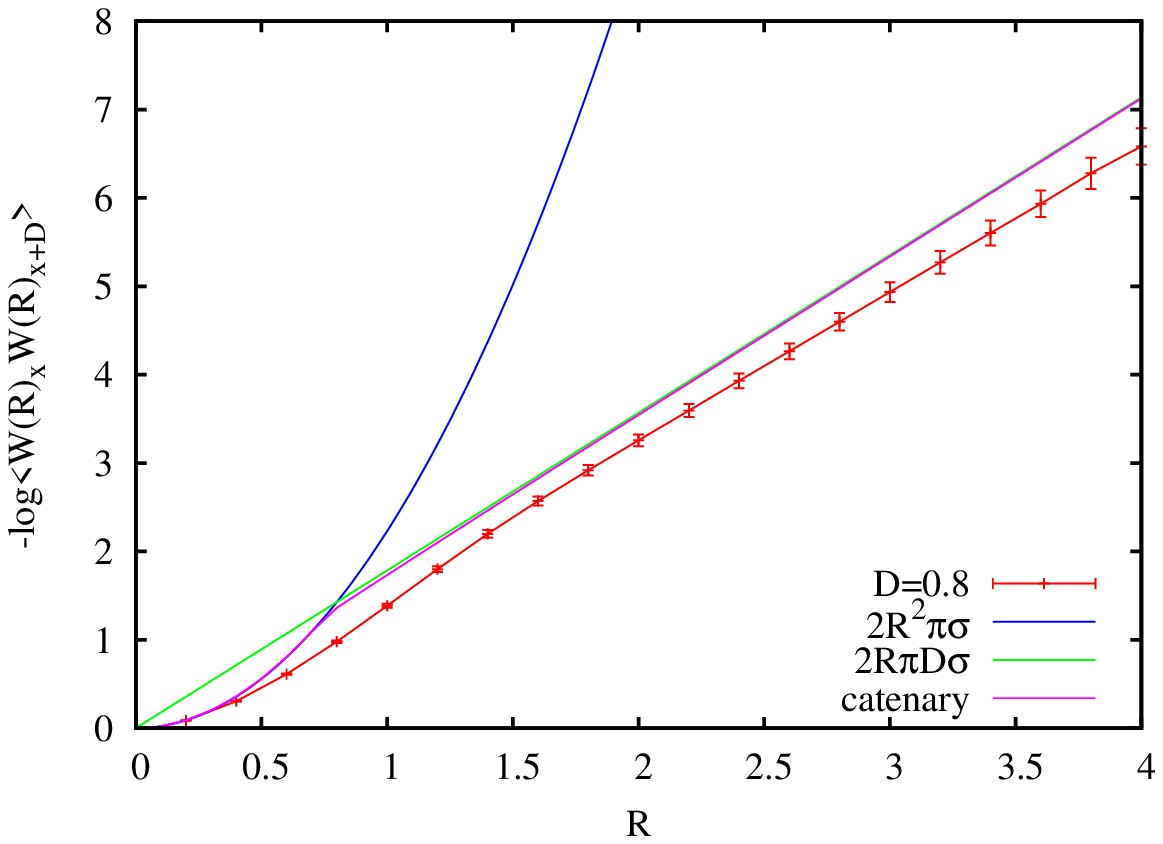}\\
	e)\includegraphics[width=.45\linewidth]{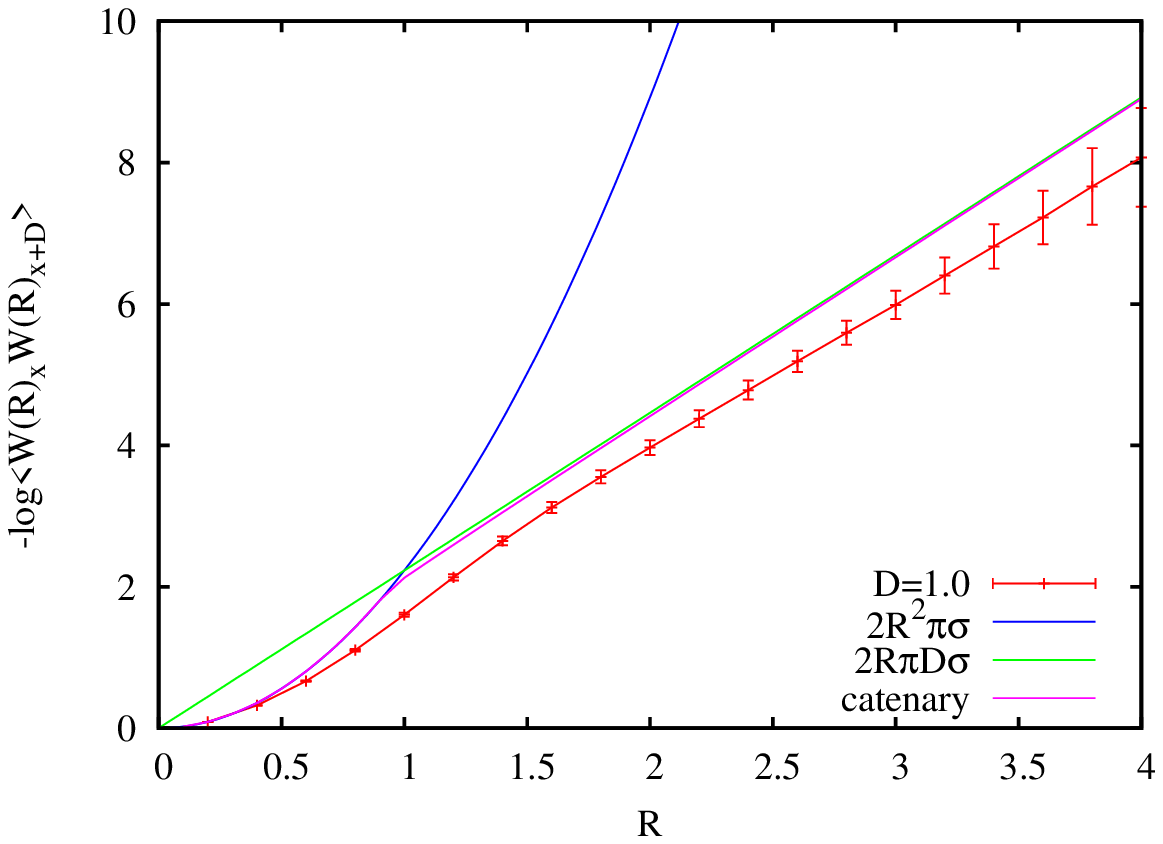}
	f)\includegraphics[width=.45\linewidth]{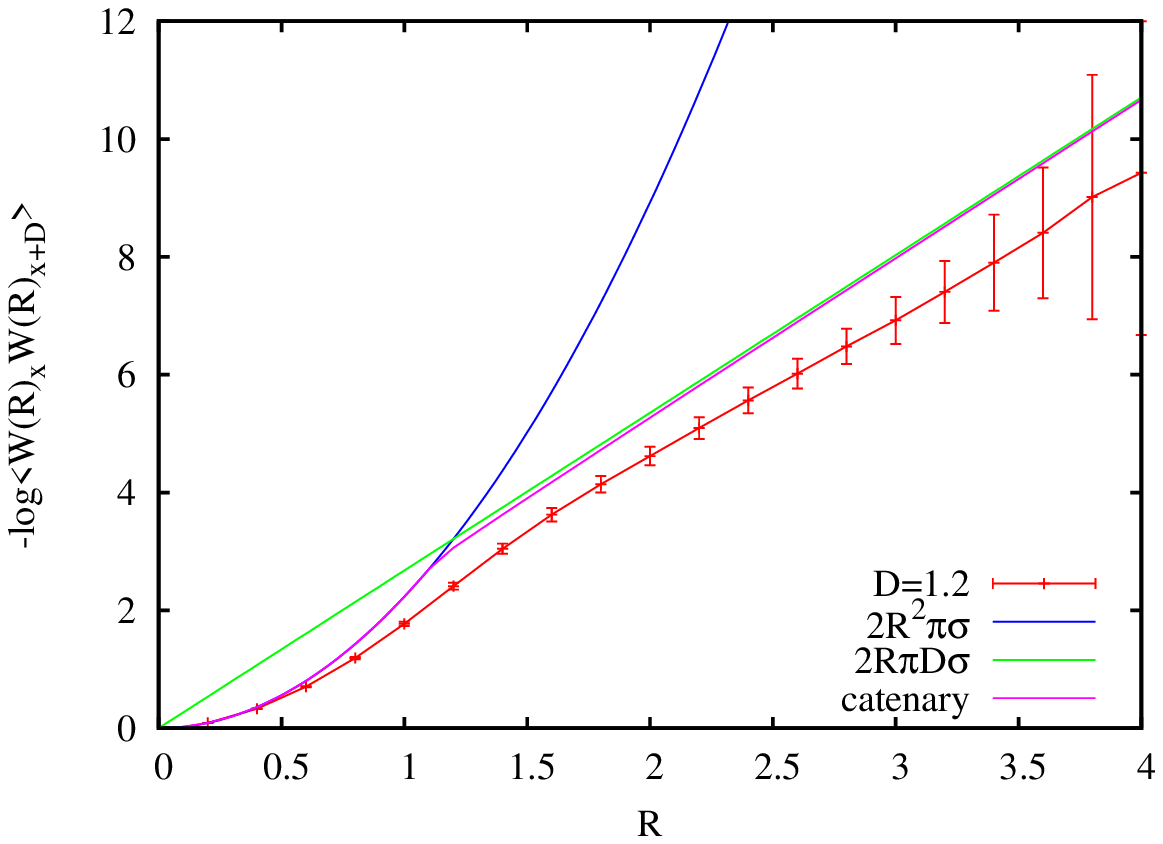}\\
	g)\includegraphics[width=.45\linewidth]{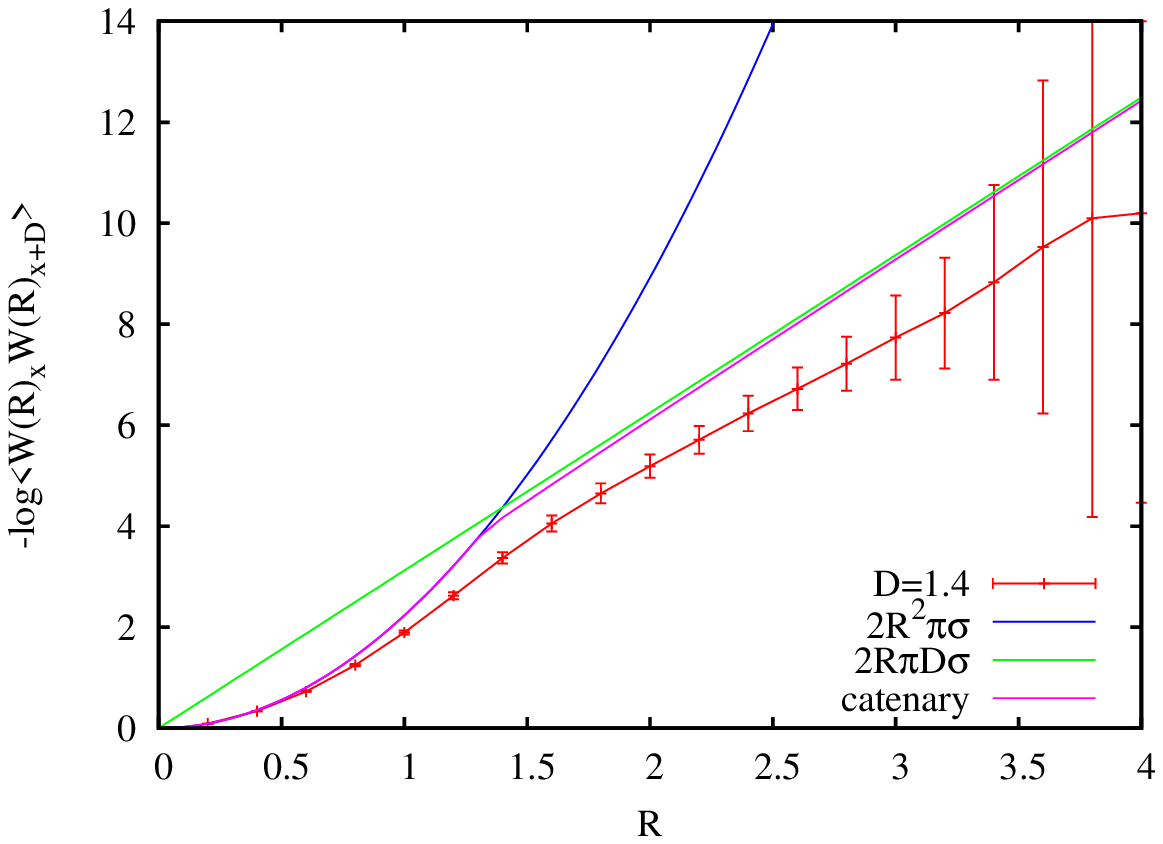}
	h)\includegraphics[width=.45\linewidth]{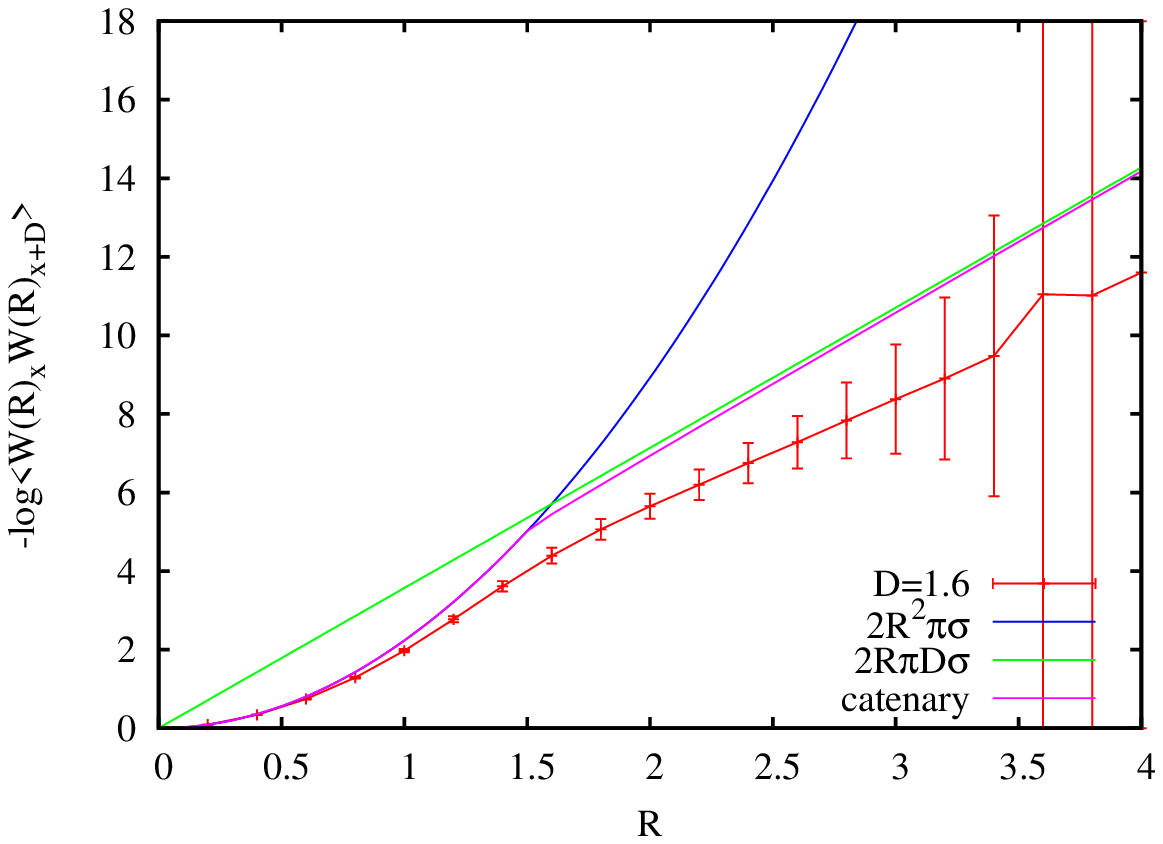}
	\caption{Circular Wilson loop correlators $-log\langle W(R)_xW(R)_{x+D}\rangle$ versus Wilson loop radii $R$ for various distances $D$ between the Wilson loops and minimal area solutions.}
	\label{fig:2dc}
\end{figure}

In the continuous model of random center vortex lines we can define circular
Wilson loops, hence we measure the correlator of two (flat) Wilson loops of
radius $R$ at a distance $D$ in $16^3$ volumes with a corresponding string
tensions $\sigma a^2=0.35$~\cite{Altarawneh:2014aa}. The Wilson loops again pick
a up a factor $(-1)$ when pierced by a vortex line. In this continuous model we
can actually go to very small distances where the data precision is still under
control, but the catenary effect is also very small. We show 3D plots of the
data and the minimal areas determined from the catenary solution
(\ref{eq:circA}) or Wilson loop sizes ($2\pi R^2$) in Fig.~\ref{fig:3dq} and 2D
cuts of the plots vs. $D$ or $R$ in Fig.~\ref{fig:2dc} and Fig.~\ref{fig:2rc}
respectively. The results show qualitative agreement with the predicted curves,
however they lie significantly below them and can not reveal the small
difference between cylinder and catenoid surface areas. 

\begin{figure}[h]
	\centering
	a)\includegraphics[width=.48\linewidth]{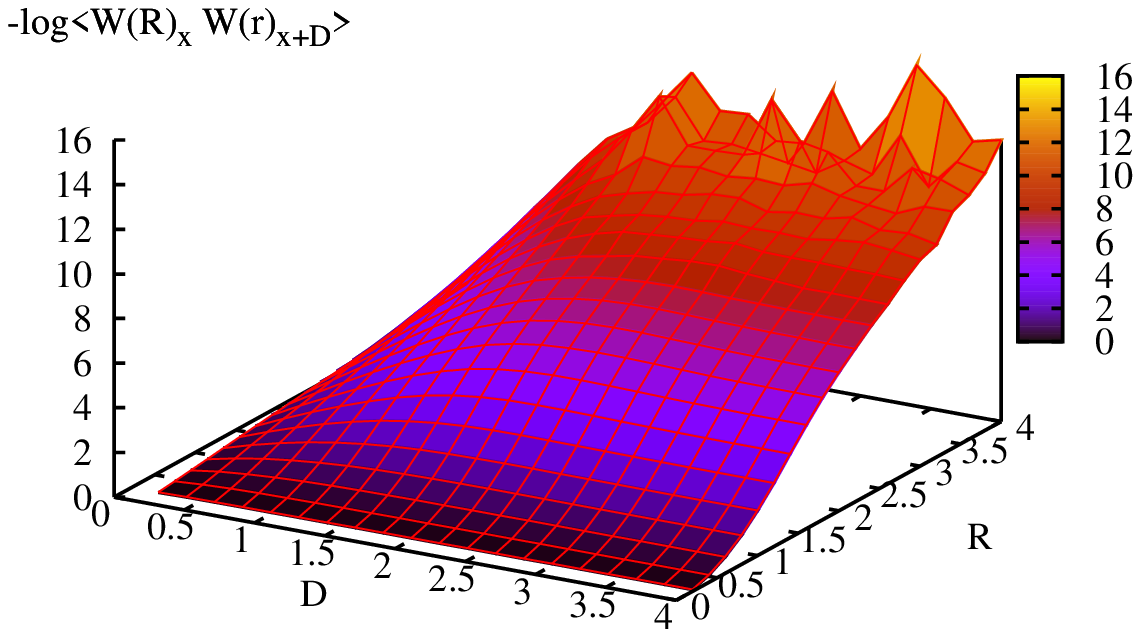}
	b)\includegraphics[width=.48\linewidth]{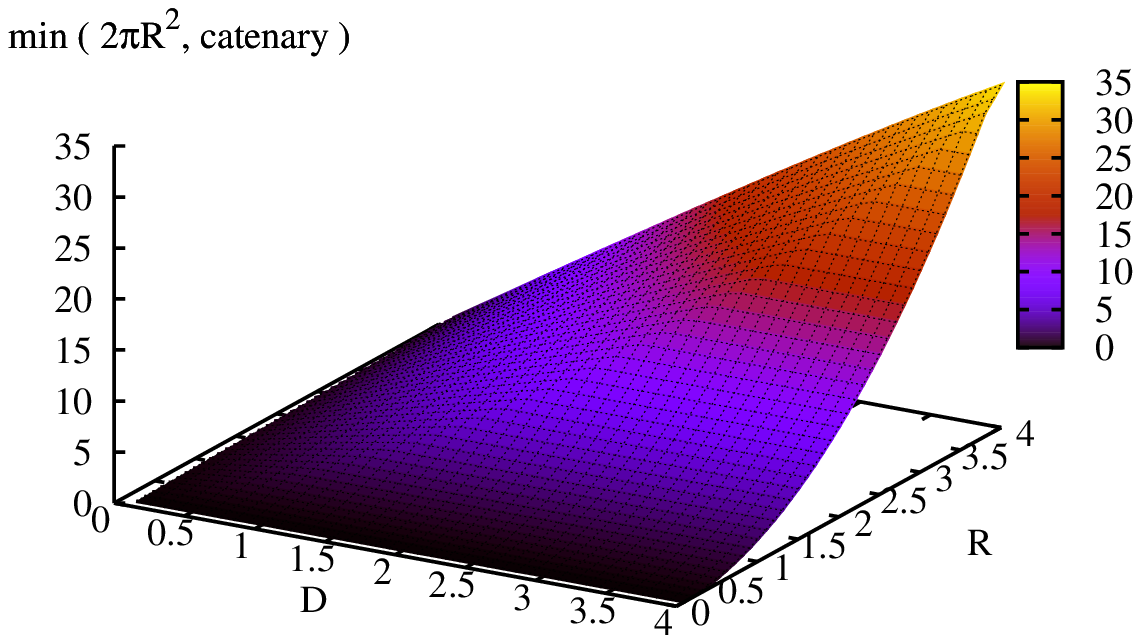}
	\caption{a) Circular Wilson loop correlators $-log\langle
	W(R)_xW(R)_{x+D}\rangle$ for various Wilson loop radii $R$ and distances $D$
	between the Wilson loops on $16^3$ lattices. b) Surface plot of the minimum
	of the two Wilson loop areas $2\pi R^2$ and the catenoid (\ref{eq:circA}).}
	\label{fig:3dc}
\end{figure}

\begin{figure}[h]
	\centering
	a)\includegraphics[width=.45\linewidth]{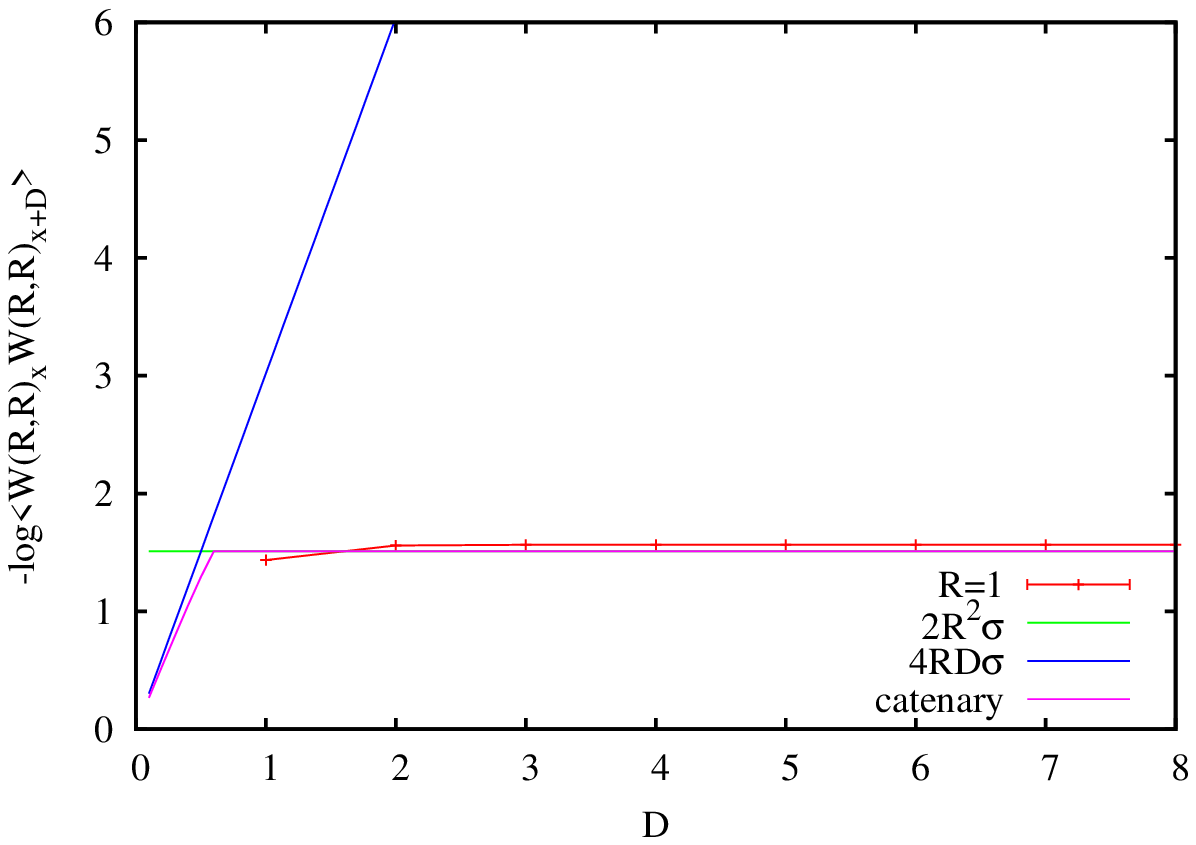}
	b)\includegraphics[width=.45\linewidth]{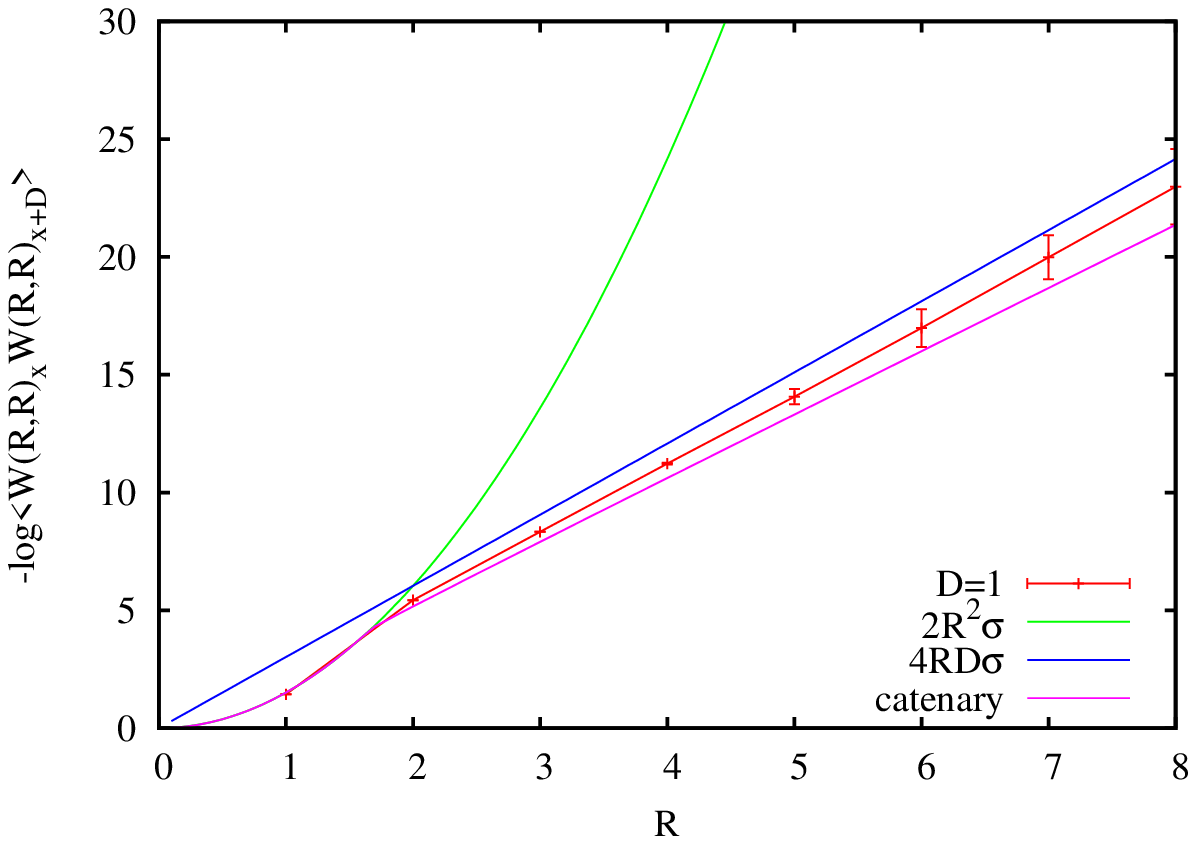}\\
	c)\includegraphics[width=.45\linewidth]{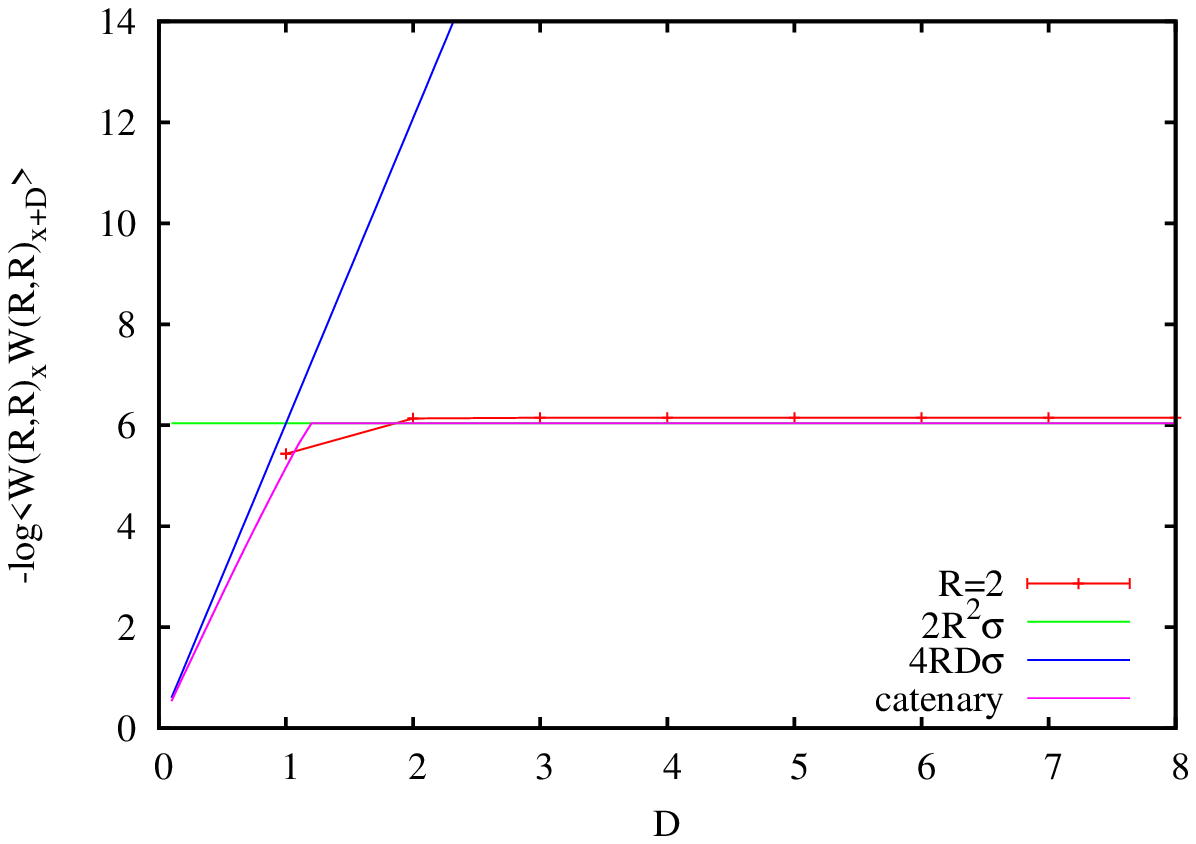}
	d)\includegraphics[width=.45\linewidth]{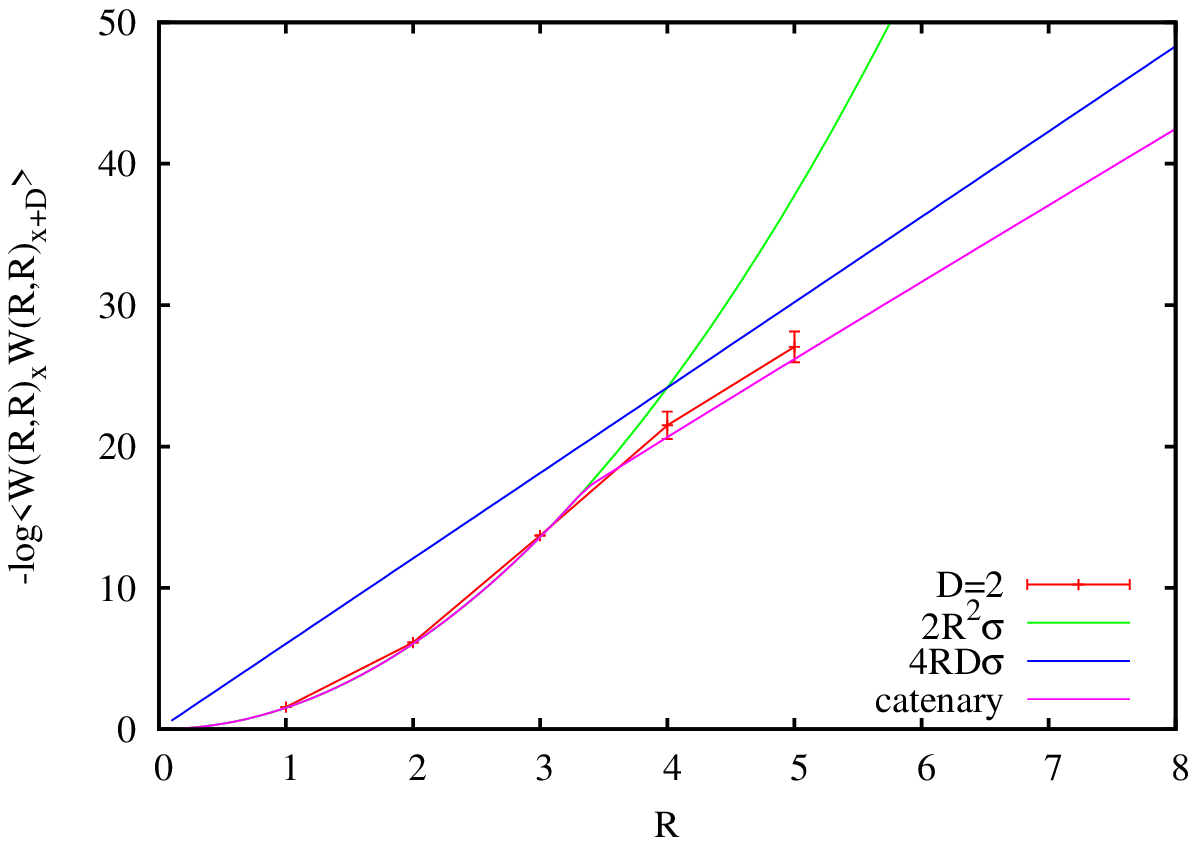}\\
	e)\includegraphics[width=.45\linewidth]{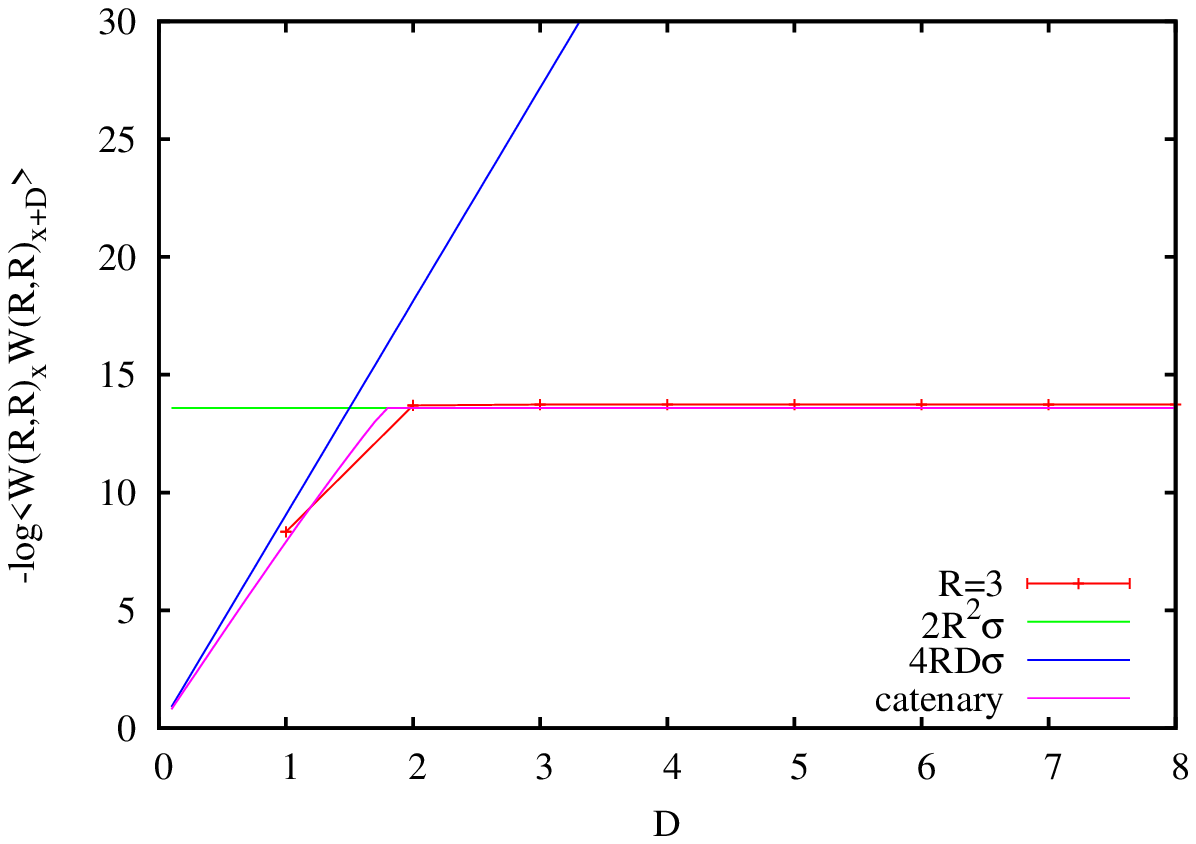}
	f)\includegraphics[width=.45\linewidth]{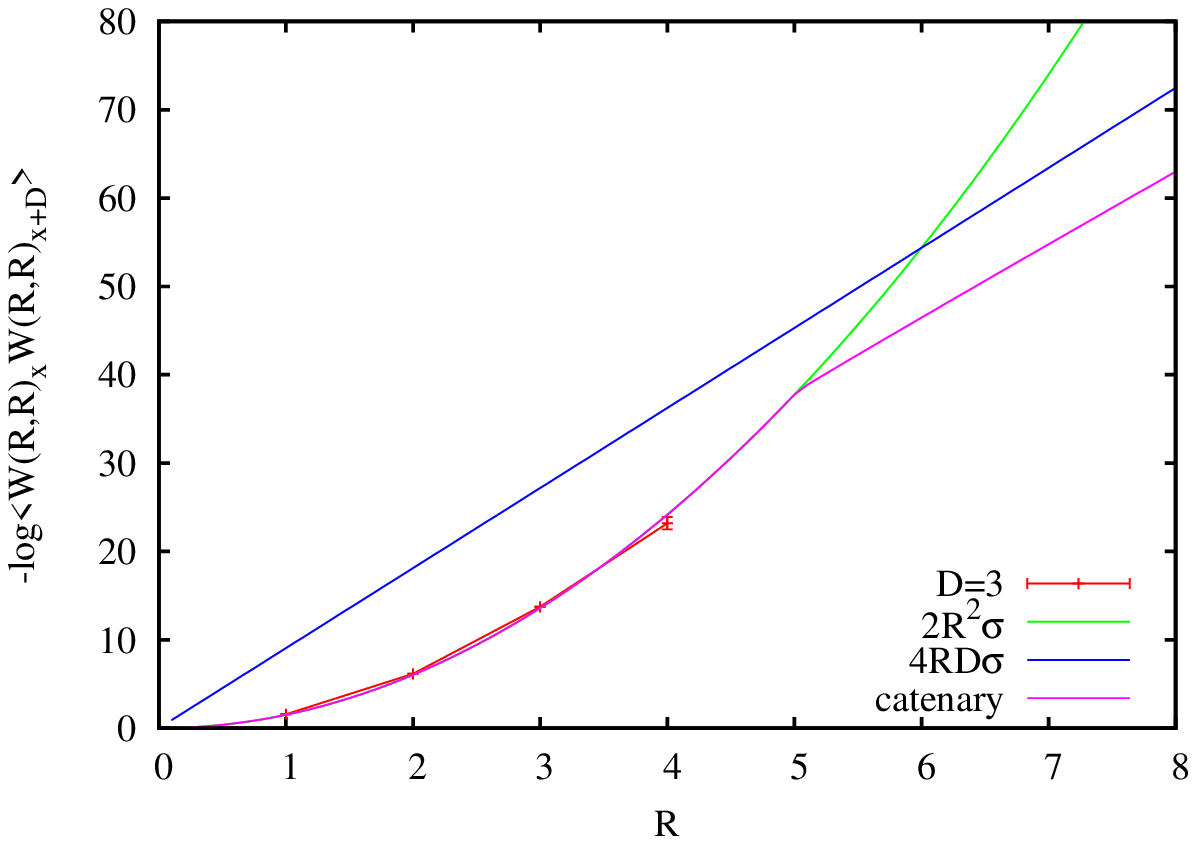}\\
	g)\includegraphics[width=.45\linewidth]{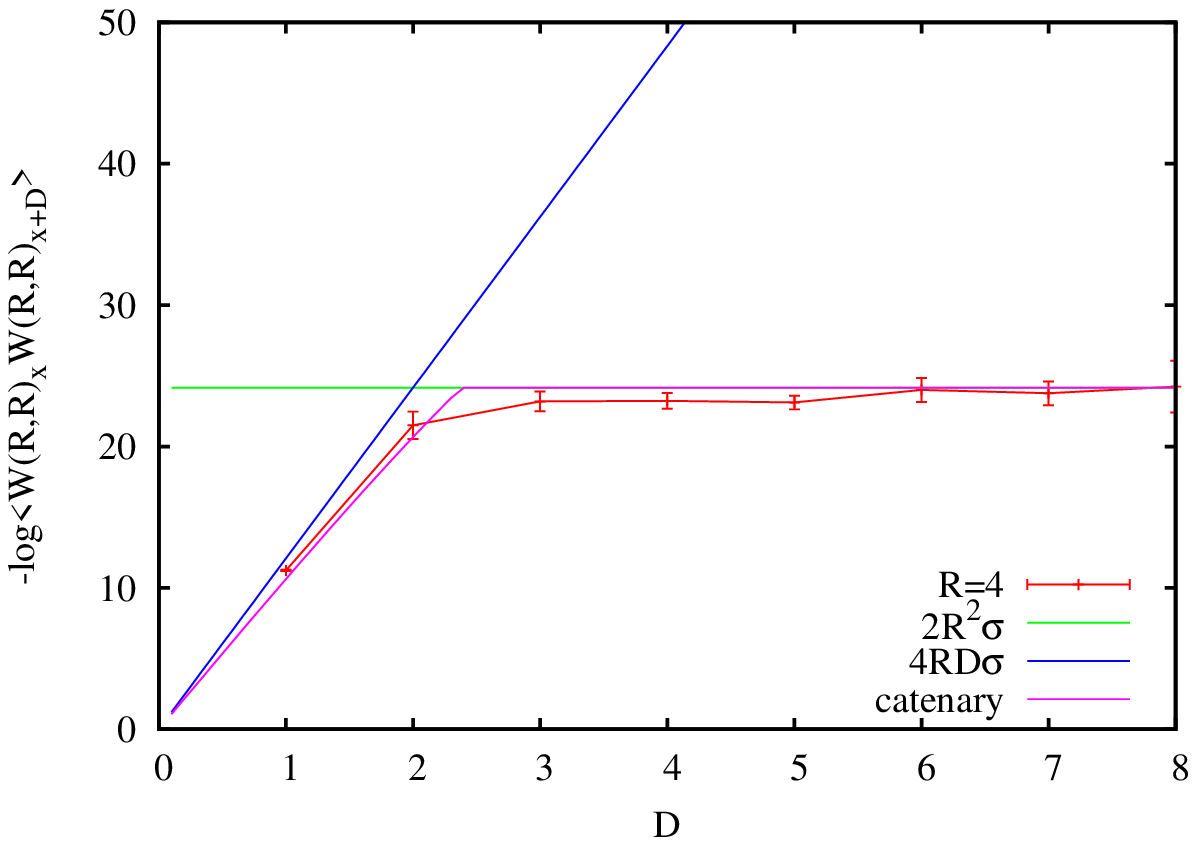}
	h)\includegraphics[width=.45\linewidth]{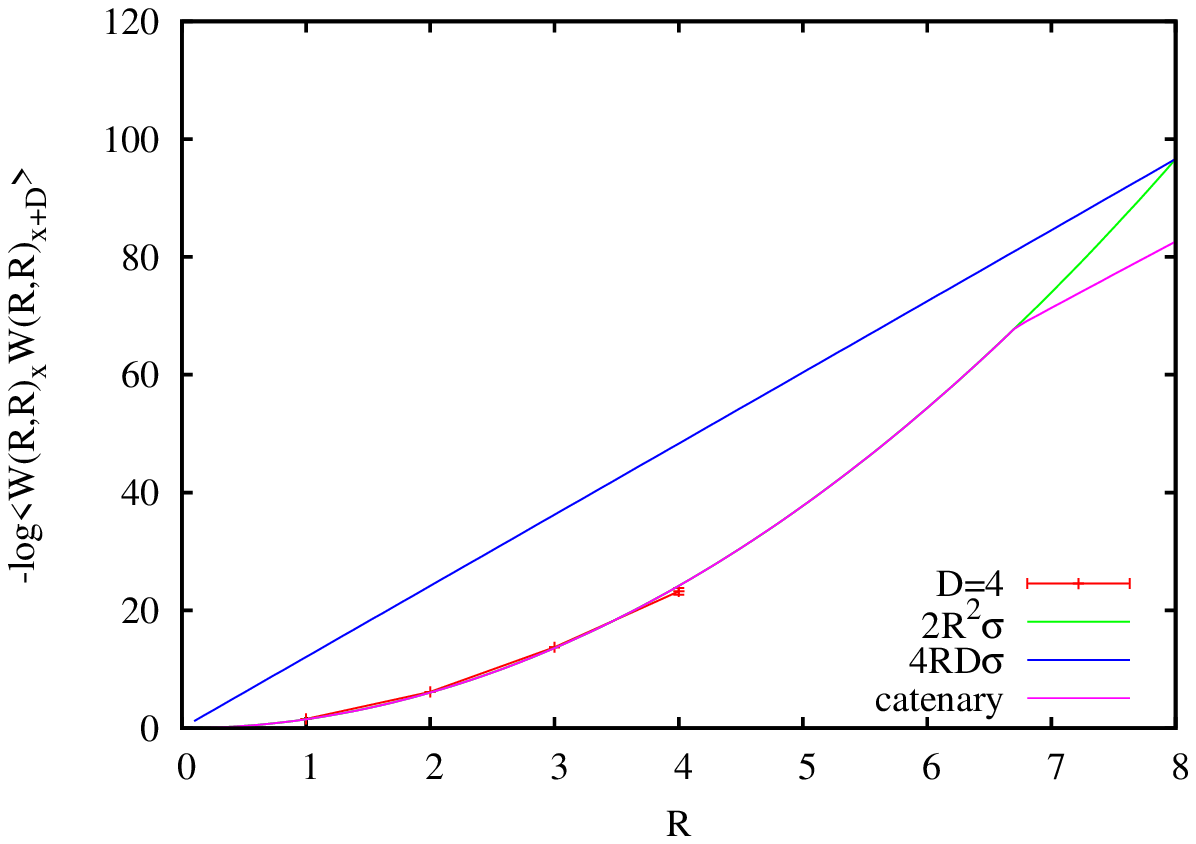}\\
	\caption{$Z(2)$ meson (quadratic Wilson loop) correlators $-log\langle
	W(R,R)_xW(R,R)_{x+D}\rangle$ versus distance $R$ between quark and
anti-quark (Wilson loop size, right column) and distance $D$ between the meson
and anti-meson (left column) on $16^4$ lattices and minimal area solutions
(catenary).}
	\label{fig:2dq}
\end{figure}

\iffalse
\begin{figure}[h]
	\centering
	a)\includegraphics[width=.45\linewidth]{WW1}
	b)\includegraphics[width=.45\linewidth]{WW2}\\
	c)\includegraphics[width=.45\linewidth]{WW3}
	d)\includegraphics[width=.45\linewidth]{WW4}\\
%e)\includegraphics[width=.45\linewidth]{WW5}
%f)\includegraphics[width=.45\linewidth]{WW6}\\
%g)\includegraphics[width=.45\linewidth]{WW7}
%h)\includegraphics[width=.45\linewidth]{WW8}
	\caption{$Z(2)$ meson (quadratic Wilson loop) correlators $-log\langle W(R,R)_xW(R,R)_{x+D}\rangle$ for various distance $R$ between quark and anti-quark (Wilson loop size) versus distance $D$ between the meson and anti-meson on $16^4$ lattices and minimal area solutions.}
	\label{fig:2dq}
\end{figure}

\begin{figure}[h]
	\centering
	a)\includegraphics[width=.45\linewidth]{WS1}
	b)\includegraphics[width=.45\linewidth]{WS2}\\
	c)\includegraphics[width=.45\linewidth]{WS3}
	d)\includegraphics[width=.45\linewidth]{WS4}\\
%e)\includegraphics[width=.45\linewidth]{WS5}
%f)\includegraphics[width=.45\linewidth]{WS6}\\
%g)\includegraphics[width=.45\linewidth]{WS7}
%h)\includegraphics[width=.45\linewidth]{WS8}
	\caption{$Z(2)$ meson (quadratic Wilson loop) correlators $-log\langle W(R,R)_xW(R,R)_{x+D}\rangle$ versus distance $R$ between quark and anti-quark (Wilson loop size), for various distances $D$ between the meson and anti-meson on $16^4$ lattices and minimal area solutions.}
	\label{fig:2rq}
\end{figure}
\fi

\clearpage

\section{Conclusions}\label{sec:con}

We measure quadratic and circular Wilson loop correlators in $Z(2)$ center
vortex models for the infrared sector of Yang-Mills theory, {\it i.e.}, a
hypercubic lattice model of random vortex surfaces and a continuous 2+1
dimensional model of random vortex lines. We further calculate the catenary
solutions for the Wilson loop configurations and the corresponding catenoid
areas. The measurements show minimal area law behavior and may indicate the
catenary effects in the hypercubic model, which physically correspond to string
surface tension leading to string constriction.

\appendix

\section{Beltrami Identity and Catenary Solution}\label{app:cat}

Starting from the Euler-Lagrange differential equation
\begin{equation}
\frac{\partial f}{\partial y}-\frac{d}{dx}(\frac{\partial f}{\partial y_x})=0\label{eq:elde}
\end{equation}
we examine the derivative of $f$ with respect to $x$
\begin{equation}
\frac{d f}{d x}=\frac{\partial f}{\partial y}y_x+\frac{\partial f}{\partial y_x}y_{xx}+\frac{\partial f}{\partial x}.
\end{equation}
Solving for the $\partial f/\partial y$ term gives
\begin{equation}
\frac{\partial f}{\partial y}y_x=\frac{\partial f}{\partial x}-\frac{\partial f}{\partial y_x}y_{xx}-\frac{d f}{d x}.\label{eq:3}
\end{equation}
Multiplying (\ref{eq:elde}) by $y_x$ and substituting the left term with the right hand side of (\ref{eq:3}) gives
\begin{equation}
\frac{d f}{d x}-\frac{\partial f}{\partial y_x}y_{xx}-\frac{\partial f}{\partial x}-y_x\frac{d}{d x}\frac{\partial f}{\partial y_x}=
-\frac{\partial f}{\partial x}+\frac{d}{d x}(f-y_x\frac{\partial f}{\partial y_x})=0.
\end{equation}
For $f_x=0$ we derive the Beltrami identity with some integration constant $a$,
\begin{equation}
\frac{d}{d x}(f-y_x\frac{\partial f}{\partial y_x})=0\quad\Rightarrow\quad f-y_x\frac{\partial f}{\partial y_x}=a,
\end{equation}
an identity in calculus of variations discovered in 1868 by Eugenio Beltrami. 
The quantity $f=y\sqrt{1+y'^2}$ from Eq.(\ref{eq:min}) in section~\ref{ssec:cwl} has in fact $f_x=0$, so we can use the Beltrami identity to obtain
\begin{equation}
y\sqrt{1+y'^2}-y'\frac{yy'}{\sqrt{1+y'^2}}=a\quad\Rightarrow\quad 
y(1+y'^2)-yy'^2=y=a\sqrt{1+y'^2}\label{eq:4}
\end{equation}
\begin{equation}
y'^2=\frac{y^2}{a^2}-1\quad\Rightarrow\quad y'=\frac{\sqrt{y^2-a^2}}{a}\quad\Rightarrow\quad \frac{dx}{dy}=\frac{1}{y'}=\frac{a}{\sqrt{y^2-a^2}}
\end{equation}
\begin{equation}
x=a\int\frac{dy}{\sqrt{y^2-a^2}}=a \cosh^{-1}(\frac{y}{a})+b\quad\Rightarrow\quad y=a\cosh(\frac{x-b}{a}),\label{eq:los}
\end{equation}
which is called a catenary, and the surface generated by rotating it around the $x$ axis is called a catenoid, see Fig.~\ref{fig:cat}.

\section{Catenary vs. Goldschmidt solution}\label{app:cvsg}

To find the maximum value of $R/D$ at which the catenary solutions (\ref{eq:los}) resp. (\ref{eq:bcs}) for the circular Wilson loop configuration can be obtained, let $p\equiv1/a$, {\it i.e.}, 
\begin{equation}
Rp=\cosh(Dp/2).\label{eq:5}
\end{equation} 
At the maximum value of $D=D^*$ (with corresponding $R^*$) it will be true that $dD/dp=0$, hence $d/dp$ of (\ref{eq:5}) is
\begin{equation}
R^*=\sinh(D^*p/2)(D^*/2+p/2\cdot dD/dp)=D^*/2\cdot\sinh(D^*p/2).\label{eq:6}
\end{equation} 
Dividing (\ref{eq:5}) at $D^*$, {\it i.e.}, $R^*p=\cosh(D^*p/2)$ by (\ref{eq:6}) yields
$D^*p/2=\coth(D^*p/2)$,
which has a solution $D^*p/2\approx1.199679$ and from (\ref{eq:6}) we derive the maximum possible value of
\begin{equation}
D/R\approx1.32549\label{eq:cmax}
\end{equation}
for catenary solutions of the circular Wilson loop configuration. For $D/R>1.32549$ only Goldschmidt solutions exist. For the quadratic Wilson loop configuration the corresponding maximum possible value of $D/R$ is given by 
\begin{equation}
D/R\approx0.88013\label{eq:qmax}.
\end{equation}

\acknowledgments{We would like to thank Michael Engelhardt for suggesting this
work and interesting discussions. This research was supported by the Erwin
Schr\"odinger Fellowship program of the Austrian Science Fund FWF (``Fonds zur
F\"orderung der wissenschaftlichen Forschung'') under Contract No. J3425-N27
(R.H.). Calculations were performed on the Phoenix and Vienna Scientific
Clusters (VSC-2 and VSC-3) at the Vienna University of Technology and the
Riddler Cluster at New Mexico State University.}

\bibliographystyle{utphys}
\bibliography{../literatur}

\end{document}